\documentclass[onecolumn,%
 aip,
 amsmath,amssymb,
 reprint,%
]{revtex4-1}

\usepackage{graphicx}
\usepackage{amsmath}
\usepackage{amsfonts}
\usepackage{amssymb}
\usepackage{hyperref}
\usepackage{dcolumn}
\usepackage{bm}
\usepackage{physics}
\usepackage{relsize}
\usepackage[utf8]{inputenc}
\usepackage[T1]{fontenc}
\usepackage{mathptmx}
\usepackage{etoolbox}

\DeclareUnicodeCharacter{2212}{-}

\makeatletter
\def\@email#1#2{%
 \endgroup
 \patchcmd{\titleblock@produce}
  {\frontmatter@RRAPformat}
  {\frontmatter@RRAPformat{\produce@RRAP{*#1\href{mailto:#2}{#2}}}\frontmatter@RRAPformat}
  {}{}
}%
\makeatother
\begin{document}

\preprint{AIP/123-QED}

\title{Symmetry-projected cluster mean-field theory applied to spin systems}

\author{Athanasios Papastathopoulos-Katsaros}
\affiliation{Department of Chemistry, Rice University, Houston, Texas 77005, USA}

\author{Thomas M. Henderson}
\affiliation{Department of Chemistry, Rice University, Houston, Texas 77005, USA}
\affiliation{Department of Physics and Astronomy, Rice University, Houston, Texas 77005, USA}

\author{Gustavo E. Scuseria}
\affiliation{Department of Chemistry, Rice University, Houston, Texas 77005, USA}
\affiliation{Department of Physics and Astronomy, Rice University, Houston, Texas 77005, USA}


\begin{abstract}
We introduce $S_z$ spin-projection based on cluster mean-field theory and apply it to the ground state of strongly-correlated spin systems. In cluster mean-field, the ground state wavefunction is written as a factorized tensor product of optimized cluster states. In previous work, we have focused on unrestricted cluster mean-field, where each cluster is  $S_z$ symmetry adapted. We here remove this restriction by introducing a generalized cluster mean-field (GcMF) theory, where each cluster is allowed to access all $S_z$ sectors, breaking $S_z$ symmetry. In addition, a projection scheme is used to restore global $S_z$, which gives rise to $S_z$ spin-projected generalized cluster mean-field (S$_z$GcMF). Both of these extensions contribute to accounting for inter-cluster correlations. We benchmark these methods on the 1D, quasi-2D, and 2D $J_1-J_2$ and $XXZ$ Heisenberg models. Our results indicate that the new methods (GcMF and S$_z$GcMF) provide a qualitative and semi-quantitative description of the Heisenberg lattices in the regimes considered, suggesting them as useful references for further inter-cluster correlations, which are discussed in this work.
\end{abstract}

\maketitle

\section{Introduction}
Physical systems can frequently be separated into constituents that are only weakly correlated with one another. Electronic structure theorists take advantage of this idea by using Hartree-Fock (HF) theory as the starting point for their calculations, where the weakly-correlated components in this example are the electrons themselves. The limitations of this approach are obvious under strong electron interactions. We can, however, also envision more complex individual constituents. For example, the individual components might be two well-separated molecules: even if the individual molecules have complex chemistry, the interactions between them when they are well-separated can be relatively straightforward to treat. Similarly, lattice models with short-range localized interactions are another prototypical example: if the interactions between sites rapidly decay with distance, the lattice sites may (but need not) separate into collections of weakly interacting "clusters" or "tiles".
\par In regular mean-field theory (HF), where the fundamental constituents are single electrons, the wavefunction can be written as a product of variationally optimized spin-orbitals. Likewise, if the fundamental constituents are collections of particles, it makes sense to imagine a wavefunction of cluster mean-field form, which is the product of variationally optimized wave functions on each cluster.
\par In a recent paper,\cite{papastathopoulos-katsaros_coupled_2022} we applied the unrestricted version of cluster-based mean-field methods (UcMF) to strongly correlated spin systems, in which each cluster was allowed to break $S^2$ but not $S_z$ symmetry. We showed that this ansatz can be very useful for the $J_1 - J_2$ Heisenberg model, and is a good starting point for other correlated methods. That work built upon our formulation of cMF for fermionic systems.\cite{jimenez-hoyos_cluster-based_2015} In the present study, we delve into two improved versions of cMF. First, we allow the individual clusters to break $S_z$ symmetry, which we call the generalized cluster mean-field (GcMF). Second, because we want the final cMF state to have good symmetry quantum numbers, we projectively restore  $S_z$ symmetry while simultaneously optimizing the cMF state. We call the ansatz $S_z$-projected generalized cluster mean-field (S$_z$GcMF), although the features that we describe can be applied to a broad spectrum of symmetries. 
\par The concept of allowing (spin) symmetries to break is not new, especially in the context of generalized Hartree-Fock (GHF).\cite{lykos_discussion_1963, lowdin_studies_1992, lefebvre_aspects_2007,valatin_generalized_1961, jimenez-hoyos_generalized_2011,hammesschiffer_advantages_1993,jake_hartreefock_2018} Restoration is similarly done with projection and it has been extensively studied in the context of HF (PHF),\cite{ring_nuclear_2004, blaizot_quantum_1986, lowdin_quantum_1955, schmid_use_2004,jimenez-hoyos_projected_2012, ghassemi_tabrizi_ground_2022, ruiz_halfprojected_2022, henderson_spin-projected_2017, garza_electronic_2014, shi_symmetry-projected_2014,cui_proper_2013} configuration interaction (CI),\cite{tsuchimochi_bridging_2017, tsuchimochi_black-box_2016, tsuchimochi_communication_2016} Møller–Plesset perturbation theory (MP2),\cite{schlegel_moeller-plesset_1988, schlegel_potential_1986, knowles_convergence_1988, knowles_projected_1988} coupled-cluster, \cite{qiuu_projected_2017, qiu_projected_2017,gomez_attenuated_2017, qiu_projected_2018, gomez_polynomial-product_2019,wahlen-strothman_merging_2017} and few determinants approximation (FED). \cite{rodriguez-guzman_variational_2014,bytautas_potential_2014, rodriguez-guzman_multireference_2013, jimenez-hoyos_multi-component_2013, rodriguez-guzman_multireference_2014, rodriguez-guzman_symmetry-projected_2012} These  projection operators, which are of the form $\int_0^{2 \pi} d\phi e^{i\phi \hat {S_z} }$, are used to restore spin symmetry by projecting out the spin component that violates symmetry and leaving only the component that has the correct symmetry.
\par In related work, wavefunction approaches that make use of cluster concepts include block-correlated coupled-cluster (BCCC), \cite{li_block-correlated_2004,wang_describing_2020}, tensor product selected configuration interaction (TPSCI), \cite{abraham_selected_2020} density matrix embedding theory (DMET), \cite{knizia_density_2012, wouters_practical_2016} active-space decomposition (ASD), \cite{parker_communication_2013} localized active space self-consistent field method (LASSCF), \cite{hermes_multiconfigurational_2019} and hierarchical mean-field theory (HMF). \cite{isaev_hierarchical_2009} However, to the best of our knowledge, some of these methods are limited by requiring all of the clusters to share the same ground state, and none of them has attempted (translational, $S^2$, and $S_z$) symmetry-broken solutions, or their restoration. Work similar to what we accomplish here has recently been presented by Ghassemi Tabrizi and Jiménez-Hoyos in Ref.~\citenum{ghassemi_tabrizi_ground_2023}. In our work, however, we  study both $XXZ$ and $J_1 - J_2$ Heisenberg lattices, focusing on $S_z$-projection at both the variation-after-projection and projection-after-variation levels, whereas Ref.~\citenum{ghassemi_tabrizi_ground_2023} deals with total spin $S$ and point-group projection on clusters. For further information on the connection between cMF and other more advanced approaches, as well as the advantages of cMF over these methods, we refer readers to Ref.~\citenum{jimenez-hoyos_cluster-based_2015} and references therein.
\par As described in our previous paper,\cite{papastathopoulos-katsaros_coupled_2022} our focus is again on spin models, specifically the $J_1 - J_2$ and $XXZ$ Heisenberg Hamiltonians. This choice is motivated by the fact that cMF is comparatively more straightforward and computationally efficient when applied to spin lattices than to more complex fermionic systems. However, it should be noted that cMF has been successfully employed in the analysis of fermionic systems, as demonstrated in prior research\cite{jimenez-hoyos_cluster-based_2015, fang_block_2007}. Lastly, we should emphasize that the primary objective of this study is to illustrate the utility of clusterization in connection with symmetry breaking and restoration, rather than significantly advancing the understanding of spin systems. To evaluate our results, we compare them to those obtained from the density matrix renormalization group (DMRG),\cite{nakatani_matrix_2018}  and exact diagonalization (FCI) for 1D and 2D finite systems.
\par The remainder of the article is structured in the following manner. Section \ref{2.0} discusses the Heisenberg models we will be exploring, and details both cMF and symmetry projection. In section \ref{3.0}, we describe some practical computational details relevant to the calculations presented in this work. The outcomes of the UcMF, GcMF, and S$_z$GcMF calculations for the 1D chain and 2D square $J_1-J_2$, and $XXZ$ Heisenberg models are presented in section \ref{4.0}. Lastly, a brief discussion of the results and some potential improvements to the approaches outlined in this paper are presented in section \ref{5.0}.

\begin{figure}
\centering
\includegraphics[scale=0.1]{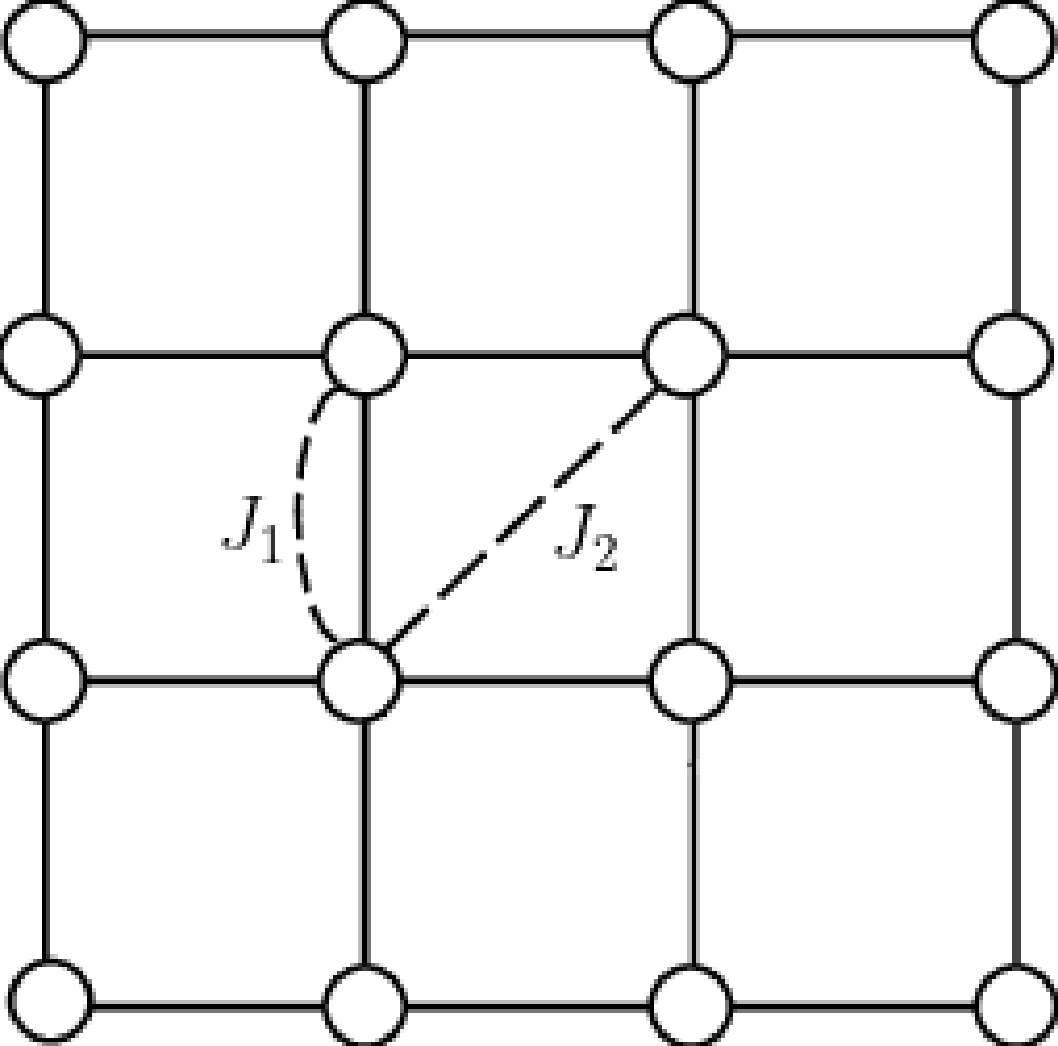}
\caption{Nearest ($J_1$) and next-nearest neighbor ($J_2$) interactions.}
\label{j1j2}
\end{figure}

\section{Formalism}\label{2.0}
\subsection{Heisenberg model}\label{2.1}
Spin lattices, particularly Heisenberg models, hold significant chemical importance. One example is the treatment of iron-sulfur clusters, like ferredoxins relevant to nitrogen fixation or photosynthesis, which have been modeled according to the Heisenberg model. \cite{chan} Another example is single-chain magnets, for example,  Cobalt(II) Thiocyanate, which has been modeled after the $XXZ$ chain\cite{rams_singlechain_2020} to study its properties. Additionally, certain electrides, conjugated hydrocarbons, and superconductors have features that mimic Heisenberg exchange interactions.\cite{apps1,apps2,apps3}
\par In this work, we examine the $XXZ$ and the $J_1$ - $J_2$ Heisenberg models. Both models describe a collection of interacting spins on a lattice of finite size $N$, but the $XXZ$ model only considers interactions between nearest-neighbors and has anisotropic interactions, which break $S^2$, while the $J_1$ - $J_2$ model includes both nearest and next-nearest neighbor interactions and has isotropic interactions where $S^2$ remains a symmetry. Both models, when the lattice is one-dimensional, are exactly solvable by Bethe ansatz.\cite{bethe_zur_1931, yang_ground-state_1966} While the 2D cases are not exactly solvable, they have been extensively studied numerically (see, for example, Refs.~\citenum{dagotto_phase_1989,schulz_finite-size_1992, richter_spin-12_2010, capriotti_spontaneous_2000, mambrini_plaquette_2006, schulz_magnetic_1996, schmalfus_quantum_2006, darradi_ground_2008, bishop_phase_1998, richter_spin-12_2015, bishop_main, jiang_spin_2012, gong_plaquette_2014, murg_exploring_2009, yu_spin-_2012, wang_constructing_2013, capriotti_resonating_2001, sandvik_finite-size_1997} for the $J_1 - J_2$ Hamiltonian and Refs.~\citenum{massaccesi_variational_2021, de_sousa_quantum_2003,cuccoli_two-dimensional_1995, jung_guide_2020, macri_bound_2021, runge_exact-diagonalization_1994,pal_colorful_2021} for the $XXZ$ model). Ref.~\citenum{pal_colorful_2021} also discusses the effect of $S_z$ projection in coloring states for the $XXZ$ model.
\par The Hamiltonian for the $XXZ$ Heisenberg model is given by:
\begin{eqnarray}
H = \sum_{\langle ij \rangle} \bigg [ \frac{1}{2} (S_i^+S_j^- + S_i^-S_j^+) + \Delta S_i^zS_j^z \bigg]
\end{eqnarray}
where $S_i^\pm$ and $S_i^z$ represent the usual spin-$\frac{1}{2}$ operators acting on site $i$, $\langle ij \rangle$ denotes nearest neighbor interactions, and $\Delta$ is the parameter that denotes the anisotropy of the model. The exact ground state of the one-dimensional case of this model has 3 distinct spin configurations:\cite{massaccesi_variational_2021} For $\Delta ~\scalebox{0.9}{\ensuremath \gtrsim}~ 1$, the magnetic correlations are Néel antiferromagnetic, for $\Delta ~ \scalebox{0.9}{\ensuremath \lesssim}~ -1$ they are ferromagnetic, and for $−1 ~ \scalebox{0.9}{\ensuremath\lesssim} ~ \Delta ~ \scalebox{0.9}{\ensuremath\lesssim} ~ 1$, the system is in the XY phase characterized by gapless excitations and long-range correlations.\cite{schollwock_quantum_2004} Moreover, it should be emphasized that at $\Delta=-1$, the ground state of the system is a maximally entangled, extreme antisymmetrized geminal power (AGP) state,\cite{zhiyuan_arxiv, massaccesi_variational_2021} with energy of $E = -N/4$, where $N$ is the number of spins. Lastly, the ferromagnetic phase can be perturbatively treated relatively easily starting from a product state (HF-like) where all spins are aligned in the z-direction. Therefore, we mostly concentrate on the more demanding region of $\Delta \geq −1$.
\par
\begin{figure}
\centering
\includegraphics[scale=0.7]{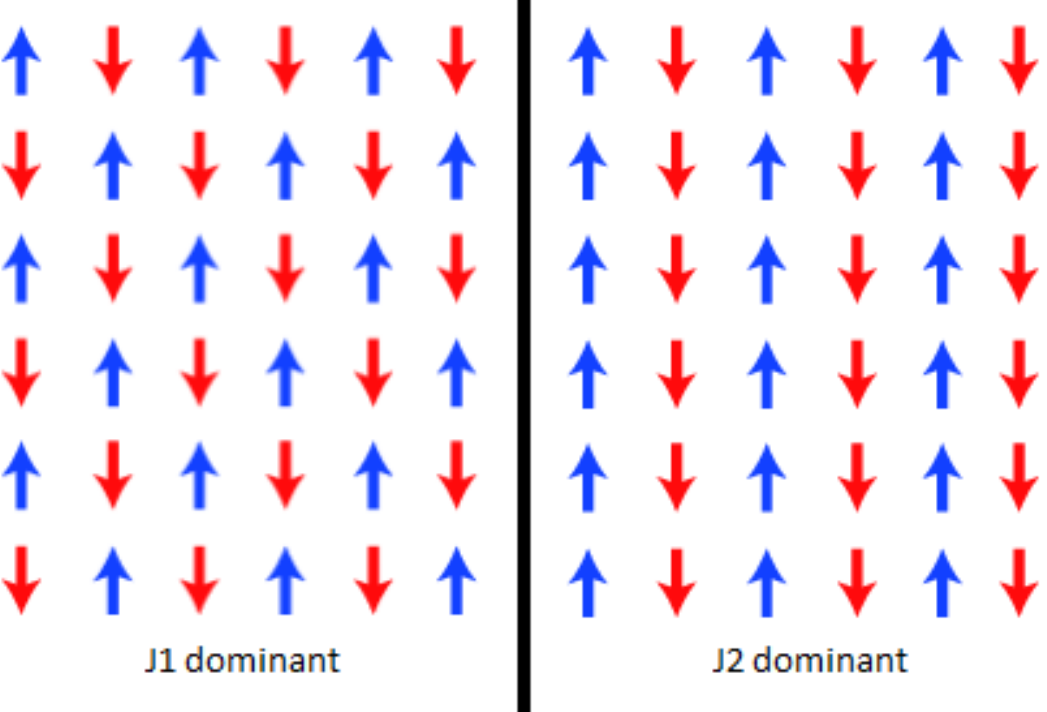}
\caption{Néel (left) and striped (right) antiferromagnetic spin configurations of the square $J_1 - J_2$ Heisenberg model.}
\label{phases}
\end{figure}

 The Hamiltonian for the $J_1$ - $J_2$ Heisenberg model can be written as:
\begin{eqnarray}
H = J_1 \sum_{\langle ij \rangle} \vec {S_i} \cdot \vec {S_j} + J_2 \sum_{\langle \langle ij \rangle \rangle} \vec {S_i} \cdot \vec {S_j}
\end{eqnarray}
where $\vec {S_i}$ is the spin-$\frac{1}{2}$ vector operator on site $i$, and $J_1$ and $J_2$ are the nearest-neighbor and next-nearest neighbor (denoted by $\langle \langle ij \rangle \rangle$) coupling coefficients, respectively (see Fig.~\ref{j1j2}). In the following, we confine ourselves to the antiferromagnetic (AFM) case $J_1, J_2 > 0$. It is important to note that for the one-dimensional case at $J_2/J_1 = 0.5$, which is called the Majumdar-Ghosh point\cite{majumdar_nextnearestneighbor_1969}, the exact ground state is UcMF of nearest-neighbor dimers (i.e., clusters of two sites). On the other hand, the two-dimensional (square) case is more complicated. This model has been studied extensively in the past two decades, using various methods. It has been established that in the regime $0 ~ \scalebox{0.9}{\ensuremath\lesssim}~ J_2/J_1 \scalebox{0.9}{\ensuremath\lesssim} ~ 0.4$, the ground state is an AFM phase with Néel order, due to the dominance of the nearest-neighbor interactions $J_1$. For $J_2/J_1 ~ \scalebox{0.9}{\ensuremath\gtrsim} ~ 0.6$, the ground state displays an AFM phase with striped long-range order character due to the dominance of the next-nearest-neighbor coupling $J_2$ (see Fig.~\ref{phases}). In the regime $0.4 ~\scalebox{0.9}{\ensuremath\lesssim} ~ J_2/J_1 \scalebox{0.9}{\ensuremath\lesssim} ~ 0.6$, which we refer to as the paramagnetic phase, the system is frustrated and the Néel and the striped orders compete. The precise nature of this intermediate ground state is still a much-debated issue, as are the type of phase transitions and the transition points (for a more extensive discussion see Refs.~\citenum{schulz_finite-size_1992, gelfand_series_1990,zhitomirsky_valence-bond_1996,takano_nonlinear_2003, isaev_hierarchical_2009, lante_ising_2006, jiang_spin_2012, wang_constructing_2013, hu_direct_2013, li_gapped_2012, nomura_dirac-type_2021, roth_high-accuracy_2022}). The two extreme configurations are illustrated schematically in Fig.~\ref{phases}. In finite systems, the same basic phenomena are observed, though of course without sharp transitions. 

\subsection{Cluster Mean-Field}
The methodology employed in this study for cluster mean-field (cMF) builds on our group's previous work in Ref.~\citenum{jimenez-hoyos_cluster-based_2015}, to which we direct the reader for a comprehensive introduction to the topic of cluster-based methods. Below, a general overview of the framework and the current extensions are presented. 
\par In regular mean-field, the wavefunction can be written as:
\begin{eqnarray}
\ket{\Phi_{HF}} = \underset{orbs}{\otimes} \ket{\phi_i}
\end{eqnarray}  where $\phi_i$ is a state defined in spin-orbital $i$. Hartree-Fock optimizes the $\phi_i$ to minimize $\bra{\Phi}H\ket{\Phi}$. A similar approach can be followed to define the cMF wavefunction:
\begin{eqnarray}
\ket{\Phi_{cMF}} = \underset{clusters}{\otimes} \ket{\phi_i}
\end{eqnarray}
where now $\phi_i$ are cluster wavefunctions chosen to minimize $\bra{\Phi}H\ket{\Phi}$. Note that if the clusters consist of single sites, cMF reduces to standard mean-field theory. 
\par While in standard mean-field theory we do not have to worry about the size and shapes of the clusters because each cluster is a single site, in the more general cMF case we obtain different results for different cluster sizes. It is therefore important to denote the clusterization scheme. For this reason, we adopt the notation cMF[m] to indicate cMF with m-site clusters in 1 dimension, cMF[$m \times n$] to indicate cMF with $m \times n$ clusters in 2 dimensions, and so forth. Although cMF permits the clusters to be of any shape and size, in this work we concentrate on dimers (clusters of 2) and tetramers (clusters of 4), as they are simpler to handle, require less computational power, and capture the short-range nature of the interactions of the spin systems here considered.
\par In addition to dependence on the size and shape of the clusters, cMF also depends on the symmetry constraints imposed on the clusters. As already mentioned, our previous investigations restricted each cluster to be an $S_z$ eigenstate. In analogy with unrestricted Hartree-Fock (UHF), we refer to this approach as unrestricted cluster mean-field (UcMF). In contrast, this work lifts that restriction and permits the clusters to be in superpositions of $S_z$ eigenstates, yielding generalized cMF (GcMF), in analogy with generalized Hartree-Fock (GHF).
\par To see the difference between UcMF and GcMF, suppose we are considering a 4-site system in which sites 1 and 2 constitute one cluster and sites 3 and 4 the other. In both UcMF and GcMF with 2-site clusters, the wavefunction is
\begin{align}
\ket{cMF[2]} = \ket{\Phi_{12}} \otimes\ket{\Phi_{34}}
\end{align}
In UcMF, when the system has global $S_z = 0$, the clusters individually do so as well, so the cluster states are
\begin{align}
\ket{\Phi_{ij}^{UcMF}} = ~c_{\uparrow \downarrow}^{ij} ~\ket{\uparrow_i \downarrow_j} ~+ ~c_{\downarrow \uparrow}^{ij} ~\ket{\downarrow_i \uparrow_j }
\end{align}
which yields
\begin{align}
\ket{UcMF[2]} = &~\ket{\Phi_{12}^{UcMF}} ~\otimes~\ket{\Phi_{34}^{UcMF}} \\ \nonumber = &~c_{\uparrow \downarrow}^{12} ~c_{\uparrow \downarrow}^{34} ~\ket{\uparrow \downarrow \uparrow \downarrow} ~+ ~c_{\uparrow \downarrow}^{12}~c_{\downarrow \uparrow}^{34}~\ket{\uparrow \downarrow \downarrow \uparrow} \\ \nonumber +& ~c_{ \downarrow \uparrow}^{12} ~c_{\uparrow \downarrow}^{34}~\ket{\downarrow \uparrow \uparrow \downarrow} ~+ ~c_{ \downarrow \uparrow}^{12} ~c_{ \downarrow \uparrow}^{34} ~\ket{\downarrow \uparrow \downarrow \uparrow}
\end{align}
Note that UcMF does not include any contributions from the configurations in which spins on sites 1 and 2 are parallel with one another. In contrast, the GcMF cluster states (for any $S_z$) are:
\begin{align}
\ket{\Phi_{ij}^{GcMF}} = & ~c_{\uparrow \uparrow}^{ij} ~\ket{\uparrow_i \uparrow_j} ~+ ~c_{\uparrow \downarrow}^{ij} ~\ket{\uparrow_i \downarrow_j } \\
\nonumber +& ~c_{\downarrow \uparrow}^{ij} ~\ket{\downarrow_i \uparrow_j} ~+ ~c_{\downarrow \downarrow}^{ij} ~\ket{\downarrow_i \downarrow_j}
\end{align}
and therefore
\begin{align}
\ket{GcMF[2]} =&~\ket{\Phi_{12}^{GcMF}} ~\otimes ~\ket{\Phi_{34}^{GcMF}} \\ \nonumber = &~\ket{\Phi_{12}^{UcMF}} ~\otimes ~\ket{\Phi_{34}^{UcMF}} \\ \nonumber ~+& ~c_{\uparrow \uparrow}^{12} ~c_{\downarrow \downarrow}^{34} ~\ket{\uparrow \uparrow \downarrow \downarrow} ~+ ~c_{\downarrow \downarrow}^{12} ~c_{\uparrow \uparrow}^{34} ~\ket{\downarrow \downarrow \uparrow \uparrow} \\ \nonumber +& ~c_{\uparrow \uparrow}^{12} ~c_{\uparrow \uparrow}^{34} ~\ket{\uparrow \uparrow \uparrow \uparrow} ~+ ~c_{\uparrow \uparrow}^{12} ~c_{\uparrow \downarrow}^{34} ~\ket{\uparrow \uparrow \uparrow \downarrow} \\ \nonumber + & ~... ~+ ~c_{\downarrow \downarrow}^{12} ~c_{\downarrow \downarrow}^{34} ~\ket{\downarrow \downarrow \downarrow \downarrow}
\end{align}
where the rest of the states are all the contributions to the GcMF wave function which break global $S_z$ symmetry. We see that GcMF includes all possible $2^4$ spin configurations, though with factorized coefficients. This provides significantly more variational freedom, though at the cost of symmetry breaking. While the foregoing discussion was restricted to the 4-site case, the basic features are generally true: GcMF includes UcMF as a special case and includes factorized coefficients on each of the $2^N$ independent spin product states in an $N$-site lattice, many of which are simply omitted in UcMF. Lastly, because GcMF has contributions from all $S_z$ sectors, directly minimizing the energy will tend to select cluster states which approximate the global minimum across all $S_z$ quantum numbers. To study an $S_z$ sector that does not correspond to the global minimum, we simply add a Lagrange multiplier to enforce that GcMF has the correct $\langle S_z \rangle$.

\subsection{The role of mean-field symmetry breaking and restoration}\label{level_of_projection}
We have seen that while the GcMF wave function includes contributions beyond those present in UcMF, some of these contributions break global $S_z$. These symmetry-breaking contributions are not present in the exact wave function (for finite systems), and one would like to eliminate them. Cluster mean-field, that is, has the same symmetry dilemma\cite{lykos_discussion_1963} as does standard mean-field theory: allowing symmetries to break affords greater variational flexibility but permits contributions from different symmetry sectors to the trial state. 
\par Similarly to what has been done extensively for GHF using projected Hartree-Fock theory,\cite{jimenez-hoyos_generalized_2011,lowdin_studies_1992} the current study goes beyond previous work by developing the projected generalized cluster mean-field method (S$_z$GcMF), which aims to restore $S_z=0$ for GcMF. $S_z$cMF[1], that is, $S_z$-projected GHF, is equivalent to our recent spin-AGP\cite{zhiyuan_arxiv} work. However, we need to emphasize that any $S_z$cMF calculation with clusters larger than 1 site, such as $S_z$cMF[2] and $S_z$cMF[$2 \times 2$], will always have more variational flexibility and provide better estimates for the energy.
\par It is important to note that this method does not require the use of mathematical projection operators in a strict sense, and instead only requires that the wavefunction $\hat P \ket{\Phi}$ is an eigenfunction of the relevant symmetry operators (in this case $\hat {S_z}$), without requiring that $\hat P$ be either Hermitian or idempotent. While quite general symmetries can be projected, we will discuss only $S_z$ projection here. For a more detailed discussion of general symmetry projection, see Refs.~\citenum{jimenez-hoyos_projected_2012, schmid_use_2004}.
\par It should be noted that in spin systems that have $S^2$ symmetry, it is possible to project not only $S_z$ but also $S^2$. Further details on this topic can be found in Appendix \ref{app1}, but the essence is that the energy can be made invariant under rotations in spin space, parameterized by three Euler angles. While $S_z$ projection requires a one-dimensional numerical integration, full spin projection involves a computational cost that is $N_{grid}^2$ times higher, where $N_{grid}$ represents the number of grid points required for one-dimensional numerical integration. In this study, our focus is primarily on $S_z$ projection due to its computational efficiency and applicability to a wider range of Hamiltonians compared to $S^2$ projection. However, we present a few selected results obtained through fully spin-projected calculations to demonstrate the significant improvement in accuracy achieved when $S^2$ projection is feasible. Lastly, the reason that we choose $S_z=0$ (and $S^2 = 0$) is that we know that the ground state of the systems that we study generally possess this quantum number, but in principle, this process can be used for any $S_z$ (and any symmetry).
\par For the case of $S_z=0$, the wavefunction that we get is
\begin{align} \label{eqn10}
\ket{S_zGcMF} &=  \hat P\ket{GcMF} \\ \nonumber &= \int_0^{2 \pi} d\phi e^{i\phi \hat {S_z} } \ket{GcMF}
\end{align}
In practice, the integral is discretized. In this work, we use $n = 2^m$ equally spaced points in a simple trapezoidal rule, so
\begin{align} \label{eqnprodform}
 \ket{S_zGcMF} &= \sum_{k=0}^{2n-1} e^{i\frac{k\pi}{n}S_z}\ket{GcMF} \\ \nonumber &= 
 \prod_{k=0}^{m-1} (1 + e^{i \frac{\pi}{2^k} \hat{S_z}})\ket{GcMF}
\end{align}
For large enough $m$, the projection becomes exact. It is easiest to understand the product form of the projector given in Eqn.~\ref{eqnprodform} by looking at which spin contaminants it eliminates. Retaining only the first term $P \sim \,1 + e^{i \pi S_z}\,$ eliminates all contaminants with odd $S_z$ and is equivalent to half-projection.\cite{ruiz_halfprojected_2022} Adding the next term in the product is equivalent to adding the points $\pi/2$ and $3\pi/2$ in the sum and eliminates contaminants with odd $S_z/2$, and so forth. A numerical comparison between the different levels of projection will be explored in sections \ref{4.2} and \ref{4.3}.
\par Lastly, we should note the distinction between the projection-after-variation (PAV) and variation-after-projection (VAP) approaches. In PAV, the unprojected state $\ket{GcMF}$ is variationally optimized in the absence of the projection operator, and the symmetry projection is carried out after the variational optimization has been completed.  In VAP, the unprojected state $\ket{GcMF}$ is variationally optimized in the presence of the projection operator. As one may expect, VAP offers a more consistent approach and greater accuracy, although at a higher computational cost. This comparison has been previously examined in the literature specifically in the context of Hartree-Fock theory.\cite{jimenez-hoyos_projected_2012, schlegel_potential_1986, mayer_behaviour_1983, koga_incorrect_1991, tsuchimochi_constrained_2011} The present work will also delve into this comparison for GcMF numerically in section \ref{4.2}. Overall, PAV can manage to reach VAP accuracy in cases where $S_z$ is severely broken, but because VAP is fully variationally optimized and PAV is not, PAV tends to be significantly less accurate in other cases. Therefore, all results in this paper have been obtained using the VAP method unless otherwise noted.

\subsection{Matrix elements and cMF optimization}
The evaluation of matrix elements for GcMF is similar to our previous work\cite{papastathopoulos-katsaros_coupled_2022} but inclusion of states with varying values of $S_z$ yields a prefactor increase. The overall scaling is unaffected. We also note that variation-after-projection (VAP) calculations are more expensive than projection-after-variation (PAV) because we must carry out symmetry projection at every iteration. However, this too does not affect the overall scaling of the method.

\section{Computational details}\label{3.0}
In this work, our calculations were performed using a locally developed code that uses the ITensor\cite{itensor} library to generate the cMF states, which makes it easy to compute the required matrix elements. We initialized the cluster states randomly and optimized the energy using the conjugate gradient\cite{atkinson_introduction_1989} algorithm with numerical gradients. For GcMF, if we wish to constrain $\langle S_z \rangle$, we use the Lagrange multiplier penalty function method. 
\par As we have noted earlier, we use dimers (2-site clusters) and tetramers (4-site clusters) in this work. These small clusters have low computational cost of cMF and serve as natural building blocks for our lattices. In 2D, our tetramers are arranged in the form of $2 \times 2$ clusters. This is because we use a square lattice, and it seems reasonable to insist that the clusters respect the shape of the lattice in order to avoid biasing the method toward one spin arrangement or another. We note that the shapes of the clusters we have chosen do not necessarily lead to the lowest cMF energy due to finite size effects. However, when the clusters are large enough, their shape is less important.
\par The UcMF calculations were performed using an equal number of up and down spins inside each cluster (only 2 states for dimers, and 6 for tetramers) to ensure that the clusters were $S_z$ eigenfunctions with $S_z = 0$. In contrast, the GcMF and S$_z$cMF calculations used all 4 states for dimers and all 16 states for tetramers. When using periodic boundary conditions (PBC), although each cluster is permitted to have different coefficients, the variational minimum in cMF tends to make each cluster identical. We note that a more accurate result could be obtained by projectively restoring lattice symmetry, but we leave this to future work. 
\par It is worth mentioning that when using GcMF and S$_z$cMF, we expect our code to converge to the lowest energy state, as long as there are no symmetries in the initial guess. This is not the case for UcMF, where different magnetic arrangements are obtained with different initial guesses; in this case, we use coefficients that have a structure reflecting that of the magnetic arrangement we wish to converge to (as was also done in Ref.~\citenum{papastathopoulos-katsaros_coupled_2022}).

\section{Results}\label{4.0}
In this section, we present the results of calculations performed using the UcMF, GcMF, and S$_z$GcMF methods on the one-dimensional (1D), quasi-one-dimensional $2 \times n$ (ladder), and two-dimensional (2D) $XXZ$ and $J_1 - J_2$ Heisenberg models discussed in section \ref{2.1}. In section \ref{4.2}, we compare our 1D results to those obtained using the density matrix renormalization group (DMRG) method. In section \ref{4.3}, we present our results for the ladder systems, and lastly, in section \ref{4.4}, we present and discuss the results of our calculations for the 2D lattices. For both quasi-one-dimensional and two-dimensional lattices, we compare our results to exact diagonalization (FCI).

\subsection{UcMF, GcMF and S$_z$GcMF results for 1D Heisenberg}\label{4.2}
\par
\begin{figure}
\centering
\includegraphics[scale=1.0]{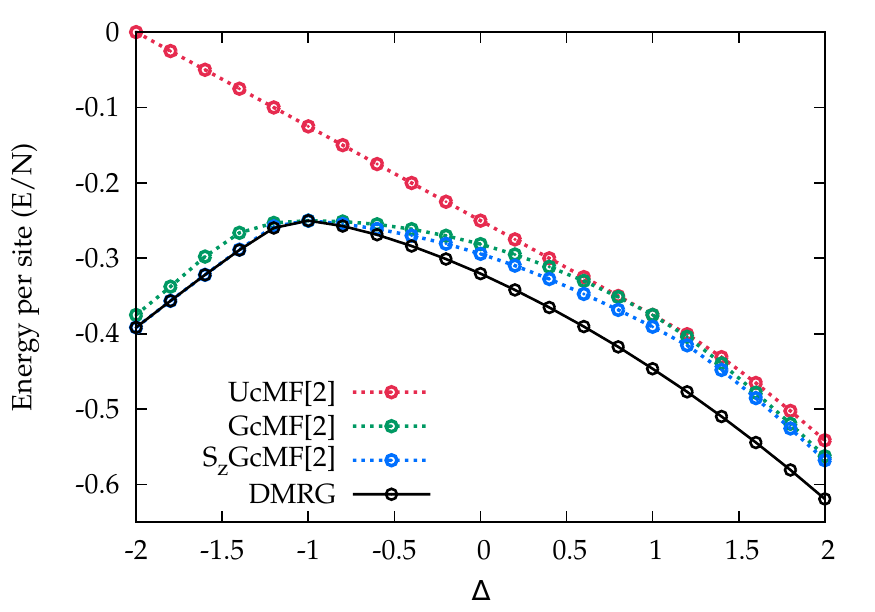}
\caption{Energy per site obtained in UcMF[2], GcMF[2], S$_z$GcMF[2] and DMRG calculations for the $XXZ$ chain as a function of $\Delta$.}
\label{udmfvsgdmf}
\end{figure}

 We begin with one-dimensional lattices, limiting ourselves to 2-site clusters. For these small clusters, the 1D $J_1 - J_2$ model does not spontaneously break $S_z$ symmetry, so UcMF[2] and GcMF[2] are equivalent. Therefore, in this subsection, we will consider only the 1D $XXZ$ model and will specialize to the 16-site $XXZ$ Hamiltonian with periodic boundary conditions. The results of our UcMF[2] and GcMF[2] calculations, along with the S$_z$GcMF[2] and exact DMRG results, are presented in Fig.~\ref{udmfvsgdmf}. 
\par Before we proceed with the presentation of our findings, it is crucial to emphasize that the global ground state for a very negative $\Delta$ is an $S_z = \pm 8$ single configuration. Consequently, due to the fact that UcMF can be defined for any $S_z$ sector, a lower UcMF[2] energy may be found. However, selecting the eigenvalue of that sector for a general Hamiltonian poses a non-trivial challenge. Therefore, in this study, we opted to utilize the $S_z = 0$ sector as it aligns most appropriately with the models under investigation and is the ground state for $\Delta > -1$. Note also that the $S_z = \pm 8$ solution is a single determinant because the corresponding sector of Hilbert space is 1-dimensional; such a simple wave function is of limited general interest.
\par Our results demonstrate that GcMF[2] yields much lower energies than UcMF[2], particularly for $\Delta$ values smaller than 0. This indicates that spin symmetry-breaking is necessary at the cMF[2] level to obtain not only a quantitatively accurate, but also a qualitatively correct solution. We should pay particular attention to the point $\Delta=-1$. Here, as we have noted, the exact ground state is an extreme AGP\cite{zhiyuan_arxiv} and is not of GcMF form.  However, all $S_z$ sectors are degenerate at this point, and GcMF even with single-site clusters (i.e., standard mean-field theory) delivers the exact energy. We note that as a consequence of this degeneracy, correlating UcMF with cluster perturbation theory\cite{papastathopoulos-katsaros_coupled_2022} by including all states of every cluster would fail since this degeneracy leads to vanishing denominators in the perturbative expansion. It is worth noting that around $\Delta=1$, the UcMF[2] and GcMF[2] results are very close to each other, which was expected since this model is equivalent to the $J_1 - J_2$ model at $J_2/J_1=0$, which also displays this property. Finally, it should be emphasized that when $\Delta \geq -1$, the global minimum is achieved at $S_z = 0$, rendering it unnecessary to impose any constraints on GcMF. However, for $\Delta < -1$, the global minimum for a system with, for instance, 16 sites corresponds to $S_z = \pm 8$, necessitating the constraint on GcMF to ensure that $\langle S_z \rangle = 0$.
\par
\begin{figure}
\centering
\includegraphics[scale=1.0]{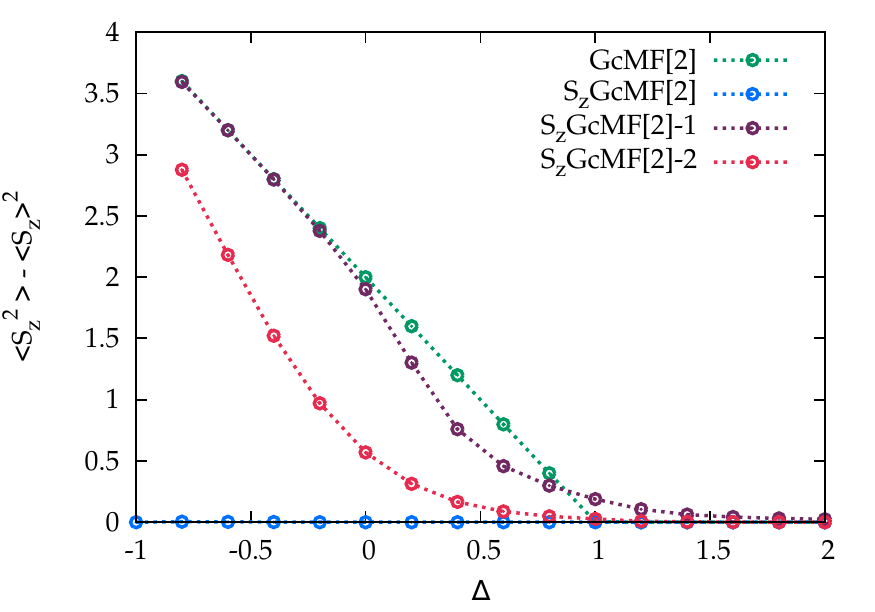}
\caption{Spin fluctuations obtained in GcMF[2] and S$_z$GcMF[4] at different levels of projection for the $XXZ$ chain as a function of $\Delta$. The number after the dash refers to the number $m$ that corresponds to the number of grid points ($n = 2^m$) used for the projection (if there is no number, full projection is assumed).}
\label{sfluc}
\end{figure}

 To evaluate the energetic improvement of projection, in addition to the calculations just discussed, we conducted S$_z$GcMF[2] calculations for the same system with identical parameters. These calculations are also summarized in Fig.~\ref{udmfvsgdmf}. There are three key observations that we wish to emphasize. Firstly, as expected, the S$_z$GcMF[2] results were overall lower in energy than the GcMF[2], and they particularly improved for values smaller than $\Delta=0$. This is most likely due to the fact that $S_z$ is more severely broken in that region, as depicted in the corresponding Fig.~\ref{sfluc}, and its restoration leads to much lower energy. Second, at $\Delta=-1$, we obtained the exact wavefunction (AGP) after the projection, which would not have been feasible in any other non-correlated way. And lastly, although this feature is only applicable for 1D $XXZ$ lattices, S$_z$GcMF[2] provides very accurate estimates of the energy for $\Delta < -1$ (the ferromagnetic region). 
\par Let us define the $S_z$ fluctuations as $\langle Sz^2 \rangle - \langle Sz \rangle ^2$. A fully projected state has zero fluctuations, and we consider the projection adequate when the fluctuations are smaller than $10^{-3}$. We note that our algorithm with random initial guesses may not converge well when we are not projecting onto the $S_z$ sector with the lowest energy (e.g. for $\Delta < -1$) and a more sophisticated implementation using a penalty term may be necessary for best performance. However, for our purposes, we did not consider it necessary to pursue further work in this direction, as it only occurs in a small part of the $XXZ$ spectrum, which is also the least interesting, as it can be easily solved by other means.\cite{massaccesi_variational_2021}
\par One advantage of the sequential half-projections we have used is that we can decide how much spin contamination to remove by simply truncating the product over $k$ in Eqn. (\ref{eqnprodform}) at a smaller value. Truncating this product is useful because it decreases the cost of the calculation, and should be done if it does not unduly compromise the quality of the projection. We examine this issue in more detail in Appendix \ref{app2}.
\par
\begin{figure}
\centering
\includegraphics[scale=1.0]{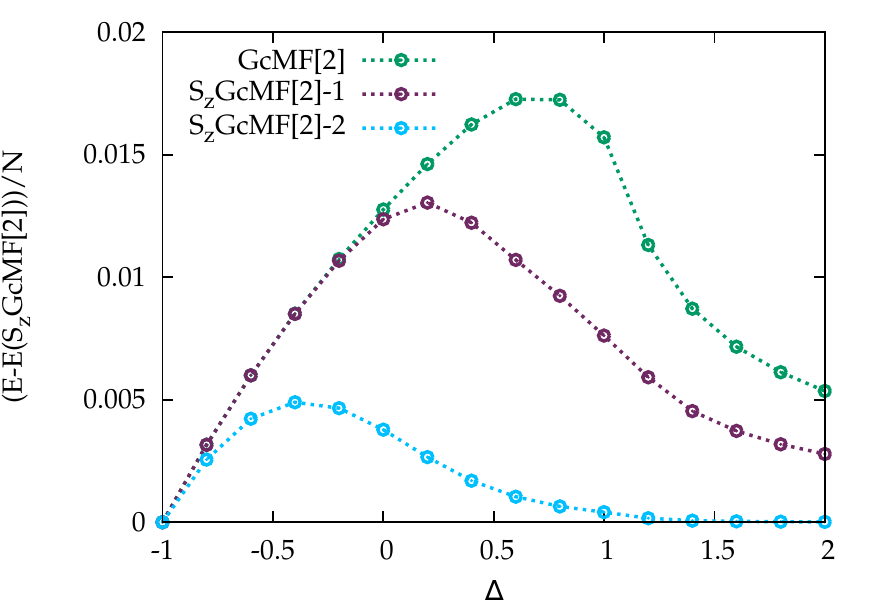}
\caption{Difference in energy per site compared to S$_z$GcMF[2] at different levels of projection for the $XXZ$ chain as a function of $\Delta$. The number after the dash refers to the number $m$ that corresponds to the number of grid points ($n = 2^m$) used for the projection.}
\label{levels}
\end{figure}

 Having addressed these, we first start by summarizing our general findings with regards to truncating the projection operator in Figs.~\ref{levels} and \ref{sfluc}. All the computational parameters are the same as previously. The number after the dash refers to the number $m$ that corresponds to the number of grid points ($n = 2^m$) used for the projection. We note two key observations. First, the number of grid points required to achieve a relatively fully projected GcMF[2] depends significantly, as expected, on the spin fluctuations of GcMF[2], which in turn are contingent on the Hamiltonian. This phenomenon is similar to what has been observed for PHF and may serve as a useful tool for pre-selecting the number of grid points when calculations are computationally intensive. Second, we focus on Fig.\ref{sfluc} around $\Delta=1$, where a somewhat counterintuitive observation emerges. Specifically, S$_z$GcMF[2]-1 and S$_z$GcMF[2]-2 exhibit greater spin fluctuations than GcMF[2], despite the former being projected. This is a consequence of the VAP approach, which tends to increase the amount of symmetry breaking in the reference GcMF state in order to lower the projected energy. This means that when the projection grid is poor, the amount of symmetry breaking may be larger than in the unprojected state, though of course the fluctuations are driven to zero by making the projection more complete.
\par
\begin{figure}
\centering
\includegraphics[scale=1.0]{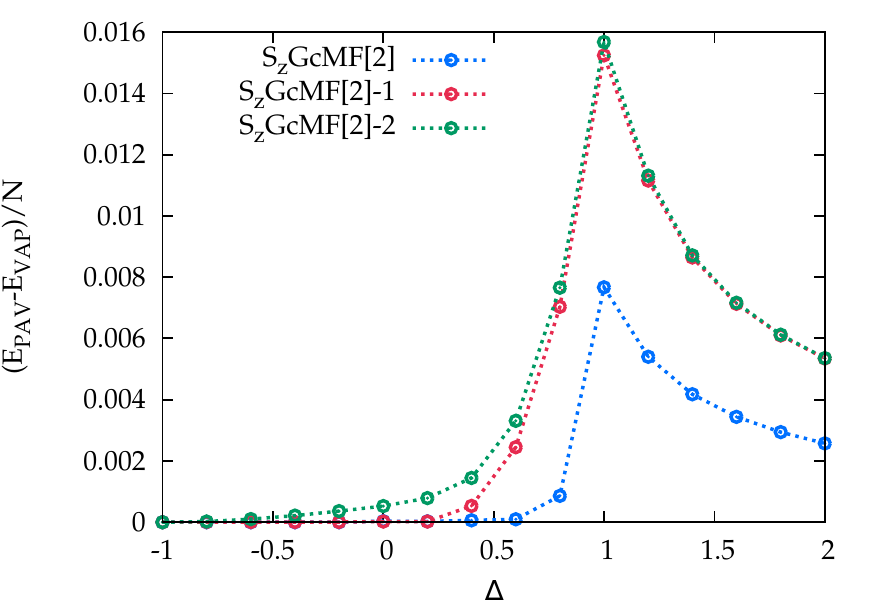}
\caption{Difference in energy per site for S$_z$GcMF[2] between VAP and PAV at different levels of projection for the $XXZ$ chain as a function of $\Delta$. The number after the dash refers to the number $m$ that corresponds to the number of grid points ($n = 2^m$) used for the projection (if there is no number, full projection is assumed).}
\label{pav}
\end{figure}

 Finally, a further characteristic of projection that warrants investigation pertains to the difference between VAP and PAV. The significance of this finding is that if the results from PAV and VAP are comparable, then the relatively more expensive and computationally intensive VAP calculation may be unnecessary and avoided in favor of PAV. To perform this analysis, we conducted additional S$_z$GcMF[2] calculations using PAV for the same system with the same parameters and varied the levels of projection. These calculations are presented in Fig.\ref{pav}, alongside the results obtained through the VAP method. We should note that these results are limited to the range $-1 \leq \Delta \leq 2$. Our analysis of Fig.~\ref{pav} revealed two distinct regions where PAV and VAP exhibit relatively similar and dissimilar behaviors, respectively. A particularly noteworthy finding is that the region where the difference between VAP and PAV is most pronounced coincides with the region where $S_z$ breaking is GcMF relatively minimal. This result aligns with our expectations since PAV can effectively restore symmetry if deemed necessary, whereas VAP's non-orthogonal configuration interaction (NOCI) character is what primarily drives energy reduction. For this reason, using PAV in the small $\Delta$ region is sufficient, which significantly decreases the cost of the projection.
 
\subsection{UcMF, GcMF and S$_z$GcMF results for quasi-1D (ladder) Heisenberg lattices}\label{4.3}
\par In this subsection, we aim to explore the potential of our theories further by performing numerical tests on the ladder $XXZ$ and $J_1 - J_2$ Heisenberg lattices. The performance of UcMF[$2 \times 2$] and GcMF[$2 \times 2$] in these quasi-1D systems is similar compared to the 1D case for the $XXZ$ Hamiltonian, and in addition, $S_z$ does not break spontaneously for the $J_1 - J_2$ model. For these reasons, we will not show any UcMF[$2 \times 2$] calculations. We will, however, explore the overall improvement that the projection offers and investigate the size-extensivity of GcMF[$2 \times 2$] and S$_z$GcMF[$2 \times 2$].
\par
\begin{figure}
\centering
\includegraphics[scale=1.0]{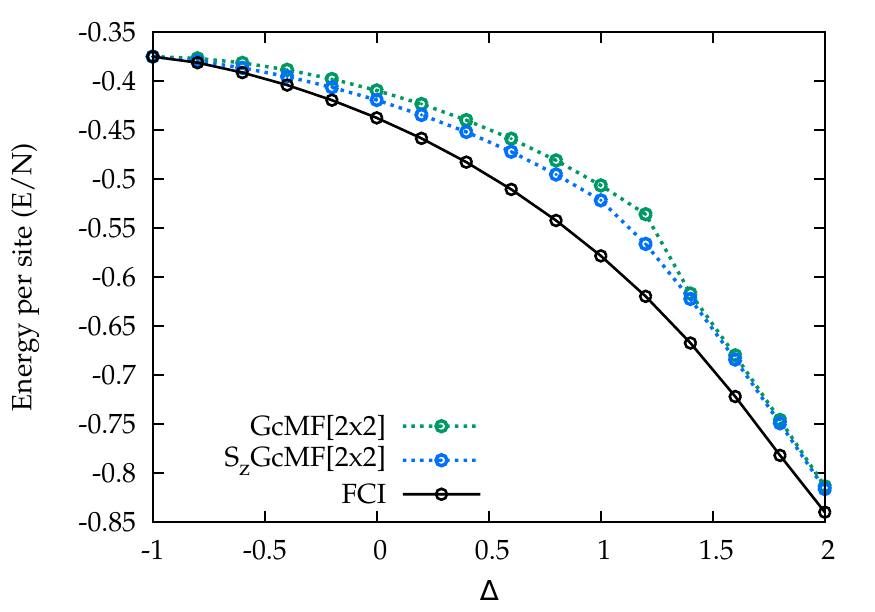}
\caption{Energy per site obtained in GcMF[$2 \times 2$], S$_z$GcMF[$2 \times 2$] and FCI calculations for the $XXZ$ ladder as a function of $\Delta$.}
\label{szgtmf_xxz}
\end{figure}

\begin{figure}
\centering
\includegraphics[scale=1.0]{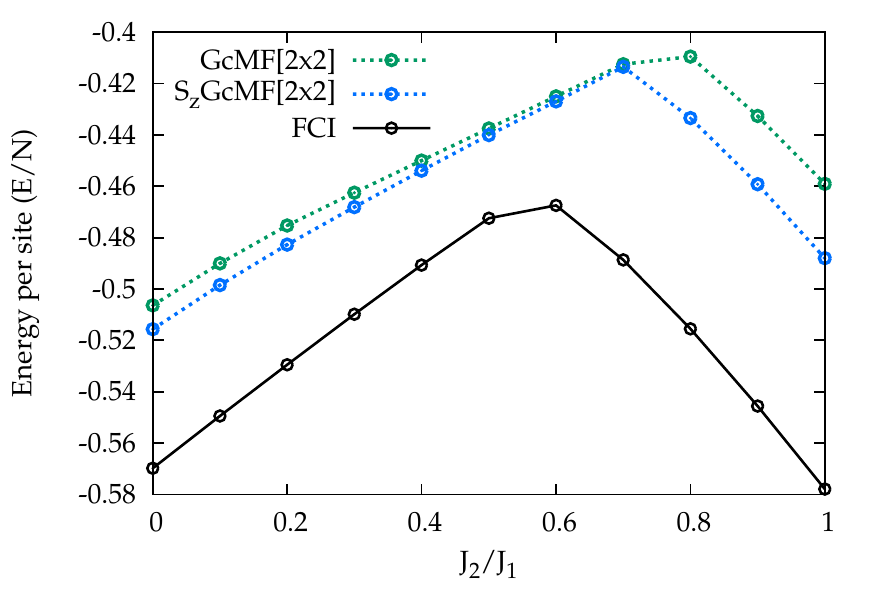}
\caption{Energy per site obtained in GcMF[$2 \times 2$], S$_z$GcMF[$2 \times 2$] and FCI calculations for the $J_1 - J_2$ ladder as a function of $J_2/J_1$.}
\label{szgtmf_j1j2}
\end{figure}

 We start by performing calculations for both models on $2 \times 12$ ladders with PBC. We can think of this calculation as six $2 \times 2$ clusters in a line. The results of our calculations, along with exact diagonalization (FCI) results, for the $XXZ$ and $J_1 - J_2$ models are presented in Fig.~\ref{szgtmf_xxz} and \ref{szgtmf_j1j2} respectively. We show $XXZ$ results only for $-1 \leq \Delta \leq 2$ for the same reasons mentioned in the previous subsection. Overall, the conclusions drawn from the previous subsection remain valid, with one exception being that S$_z$GcMF[$2 \times 2$] does not yield as precise energy estimates for $\Delta < -1$ (the ferromagnetic region) as it did before. We do not show or discuss this data here but will examine this shortcoming further in the subsequent subsection, which focuses on genuinely 2D systems. Nonetheless, two crucial observations still deserve attention. First, although $S_z$ is not explicitly broken by the $J_1 - J_2$ Hamiltonian at the GcMF[$2 \times 2$] level, the energy can be greatly improved by utilizing S$_z$GcMF[$2 \times 2$]. This is particularly more pronounced for the striped antiferromagnetic phase but holds for all the spectrum. As previously discussed, this is mainly attributed to the fact that the projection, implemented via VAP, can be thought of as a linear combination of $S_z$-broken GcMF[$2 \times 2$] wavefunctions, which may only break $S_z$ in the presence of each other and is responsible for energy reduction. Secondly, for both Hamiltonians, S$_z$GcMF[$2 \times 2$] results in smoother curves, leading to more accurate qualitative (and quantitative) descriptions, in comparison with GcMF[$2 \times 2$]. This observation is critical since the ultimate objective of this paper is to develop a more advanced ansatz that can improve not only the energy estimates but also other qualitative features, which will allow correlated methods to work with less effort. However, it should be pointed out that the smooth character of the FCI solution is due to finite-size effects, and in the thermodynamic limit there are genuine phase transitions.
\par Regarding the size-extensivity of the projected wavefunctions, it is well-known that projected methods are typically not size-extensive, meaning that they are unable to provide accurate results when studying thermodynamic limit properties (as discussed in Ref.~\citenum{jimenez-hoyos_projected_2012}). More precisely, the size-extensive component of a projected wave function is the same as that of the unprojected state, so that in the thermodynamic limit $S_z$GcMF and GcMF should yield identical energies per cluster. However, for our purposes, this may not necessarily be an issue. Firstly, real chemical systems are finite, and thus projected methods can still provide valuable insights into energy properties. Additionally, projected methods can accurately predict quantum numbers even if the energy estimates are not very accurate. For these reasons, we aim to explore the size-extensivity of the GcMF methods in our study.
\par
\begin{figure}
\centering
\includegraphics[scale=1.0]{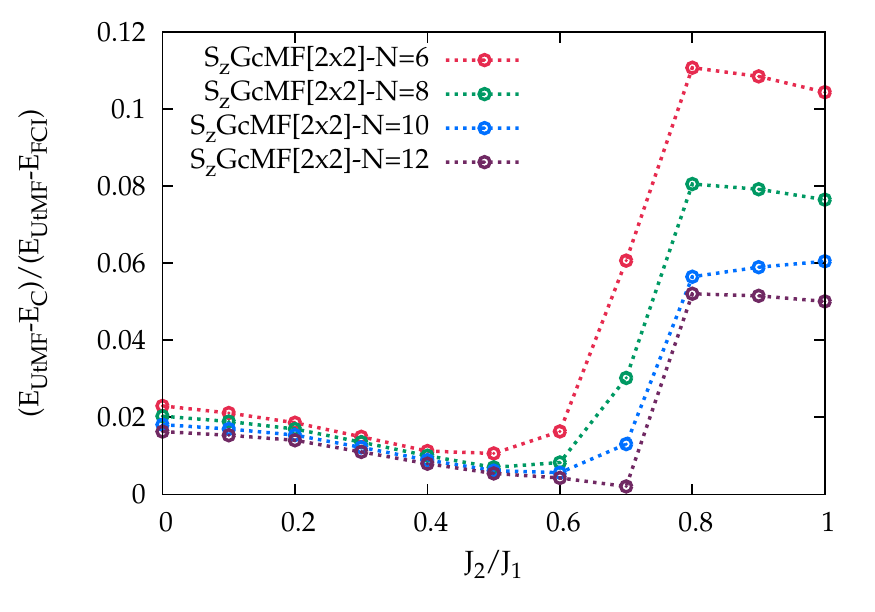}
\caption{Portion of additional energy recovered by GcMF[$2 \times 2$] and S$_z$GcMF[$2 \times 2$] compared to UcMF[$2 \times 2$], for the $J_1 - J_2$ model as a function of $J_2/J_1$ using $2 \times N=6,8,10,12$ ladders.}
\label{j1j2_system_size}
\end{figure}

 To achieve this, we performed calculations for the $J_1 - J_2$ Hamiltonian using $2 \times N=6,8,10,12$ ladders with PBC, and our results are presented in Fig.~\ref{j1j2_system_size}. We observed two distinct cases even though spontaneous $S_z$ breaking never occurs; the behavior of $S_z$GcMF[$2 \times 2$] significantly changes at $J_2/J_1 \approx 0.7$ as the method is significantly more accurate for the striped antiferromagnetic case.

The important and interesting observation was that the significance of the projection decreases as the system size increases. However, the extent to which this occurs varies depending on the phase, which, in turn, depends on the Hamiltonian. For instance, the diminishing quality of the projection is much more pronounced for the striped antiferromagnetic region of $J_1 - J_2$ compared to the rest of the spectrum. However, there is no clear explanation for why and when this occurs, as it requires an extensive study of the exact solution of the Hamiltonian. 

\subsection{UcMF, GcMF and S$_z$GcMF results for 2D Heisenberg}\label{4.4}
\begin{figure}
\centering
\includegraphics[scale=1.0]{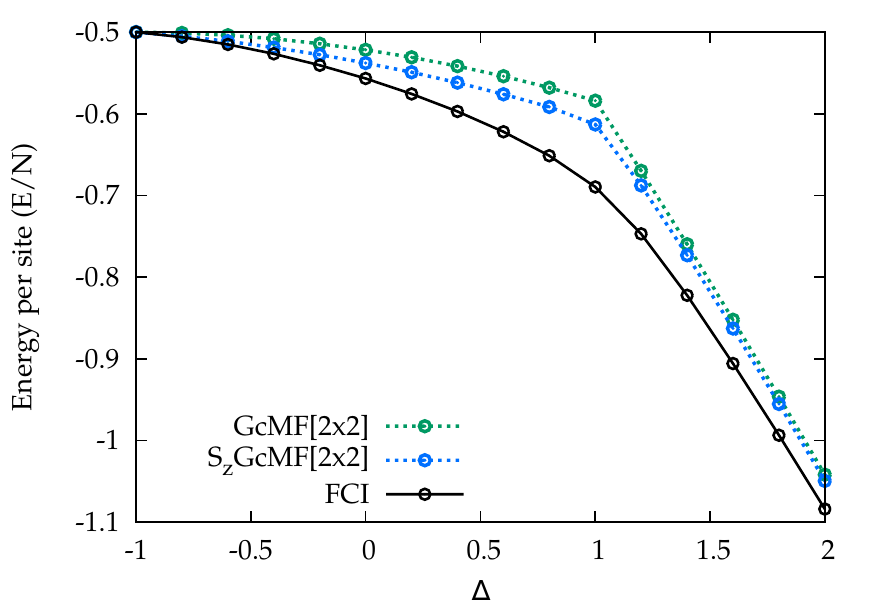}
\caption{Energy per site obtained in GcMF[$2 \times 2$] and S$_z$GcMF[$2 \times 2$] for the 2D $XXZ$ Heisenberg lattice.}
\label{2d_xxz}
\end{figure}

\begin{figure}
\centering
\includegraphics[scale=1.0]{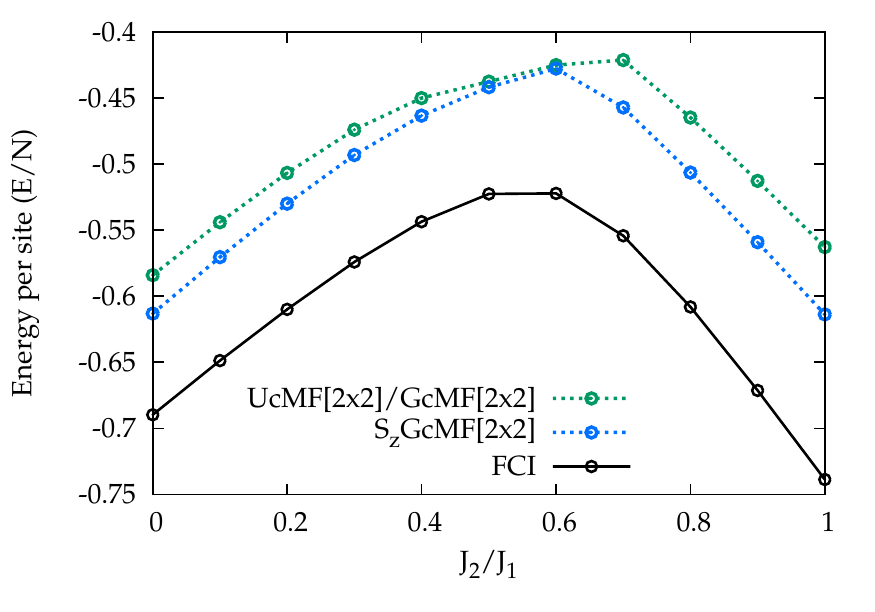}
\caption{Energy per site obtained in GcMF[$2 \times 2$] and S$_z$GcMF[$2 \times 2$] for the 2D $J_1 - J_2$ Heisenberg lattice.}
\label{2d_j1j2}
\end{figure}

In the final subsection of our results, we aim to explore the accuracy of GcMF[$2 \times 2$] and $S_z$GcMF[$2 \times 2$] as viable ansatze even in the context of two-dimensional (2D) models. This is achieved by presenting our findings for the $4 \times 6$ $J_1 - J_2$ and $XXZ$ Heisenberg models and comparing them with exact diagonalization (FCI) results (see Figs.~\ref{2d_xxz} and \ref{2d_j1j2}). Our previous observations still hold true, as both methodologies yield accurate estimates of the energy for the $XXZ$ model, particularly within the small $\Delta$ region. In addition, the largest discrepancy is noted at $\Delta \approx 1$ (which corresponds to $J_2/J_1 \approx 0$). In the $J_1 - J_2$ model, $S_z$GcMF results in a curve which, in contrast to those from GcMF or UcMF, has a similar shape to the exact one and seems to have similar accuracy for the three different magnetic structures (the Néel antiferromagnet, the paramagnet, and the striped antiferromagnet).

\section{Discussion}\label{5.0}
In section \ref{2.0}, we discussed the cluster mean-field method for addressing strongly-correlated spin systems. We have here further developed this approach by introducing the generalized cluster mean-field formalism, which allows for $S_z$ breaking within each cluster, and subsequently, the S$_z$GcMF projection formalism. Both methods partially account for the inter-cluster correlations that are absent in UcMF. These extensions can be employed as variational ansatze for the ground state wavefunction and have been shown to yield more accurate variational estimates than the UcMF approach. In sections \ref{4.2}, \ref{4.3} and \ref{4.4}, our results demonstrate that the cluster-based approaches, especially GcMF and S$_z$GcMF, can quantitatively capture very well the ground state physics of the benchmark models, especially the 1D, ladder, and 2D square $XXZ$ Heisenberg model, but can also provide a qualitatively accurate ansatz for the $J_1 - J_2$ Heisenberg models. In the same sections, we have also discussed the size-extensivity of these methods and the different levels of projection in terms of grid points used, as well as compared the variation-after-projection (VAP) with the projection-after-variation (PAV) schemes. These results further provide evidence that these cluster-based wavefunctions can be used for qualitatively describing spin lattices in general, and could serve as references for more advanced correlated methods.
\par In previous work, we have discussed correlating UcMF,\cite{jimenez-hoyos_cluster-based_2015, papastathopoulos-katsaros_coupled_2022} and all of these ideas can straightforwardly be applied to GcMF as well. Loosely, these approaches rely on the analogy between cMF and Hartree-Fock to introduce cluster-based versions of perturbation theory or coupled-cluster theory. One could anticipate that due to the breaking of the $S_z$ symmetry by GcMF, the process of correlating it becomes notably more challenging. However, in practical applications, the calculation of matrix elements between separate cluster states remains straightforward, albeit with an increased number of matrix elements that need to be computed. On the other hand, correlating the projected state is more complicated. Considerable effort has been devoted to a correlate-then-project approach in which one correlates the broken-symmetry mean-field with standard methods, then symmetry-projects the resulting wave function.\cite{tsuchimochi_orbital-invariant_2018, duguet_symmetry_2015, song_power_2022, qiu_projected_2017} These ideas could in principle be used with cMF, although they are technically rather challenging. Alternatively, we could employ Jastrow-like operators, which have been attempted for the antisymmetrized geminal power (AGP) wavefunction \cite{khamoshi_exploring_2021, henderson_correlating_2020} and have demonstrated promising results. Incorporating such operators may further improve the accuracy of our cluster-based wavefunctions and enable their application to more complex systems. Lastly, we can use a similar approach to the few determinant approximation (FED), \cite{rodriguez-guzman_variational_2014,bytautas_potential_2014, rodriguez-guzman_multireference_2013, jimenez-hoyos_multi-component_2013, rodriguez-guzman_multireference_2014, rodriguez-guzman_symmetry-projected_2012} but instead of determinants we can utilize cluster product states and optimize them in a NOCI manner.
\par Our results have been presented only for spin lattices, but they show that cMF is promising for these demanding model Hamiltonians. We emphasize that the same basic techniques we have described here can be used in general chemical systems. One could envision, for example, choosing the clusters as different functional groups in an organic molecule. GcMF can be especially useful when it comes to chemical systems, where the cluster $S_z$ eigenvalues cannot be predetermined. Another occasion is when there are external magnetic fields involved, which can alter $S_z$ significantly, and require such a general approach. The results presented here suggest the promise of these kinds of techniques and also imply that the best accuracy requires permitting each cluster to break any or even all symmetries of the global Hamiltonian.


\begin{acknowledgements}
This work was supported by the U.S. Department of Energy, Office of Basic Energy Sciences, Computational and Theoretical Chemistry Program under Award No. DE-FG02-09ER16053. G.E.S. acknowledges support as a Welch Foundation Chair (Grant No. C-0036).
\end{acknowledgements}

\section*{data availability}
The data that support the findings of this study are available from the corresponding author upon reasonable request.

\appendix
\section{$S^2$ projection}\label{app1}
Here we provide a brief overview of $S^2$ projection. For a more detailed discussion,  see Refs.~\citenum{jimenez-hoyos_projected_2012, schmid_use_2004}.
\par To restore $S^2$, the projection has to make the energy invariant under rotations in spin space, parameterized by three Euler angles. Contrary to $S_z$, where only the one-dimensional numerical integration was required, for $S^2$ a three-dimensional numerical integration is required, because of the three Euler angles. For the case of $S^2=0$, similarly to eq.~\ref{eqn10}, we can write the wavefunction as
\begin{align} 
\ket{S^2GcMF} &=  \hat P\ket{GcMF} \\ \nonumber &= \int_0^{2 \pi} d\alpha e^{i\alpha \hat {S_z} } \int_0^{\pi} sin(\beta)d\beta e^{i\beta \hat {S_y} } \int_0^{2 \pi} d\gamma e^{i\gamma \hat {S_z} } \ket{GcMF}
\end{align}
where $\alpha, \beta$ and $\gamma$ are the three Euler angles. Likewise, as for eq.~\ref{eqnprodform}, in practice, the integral is discretized, and for large enough $m$, the projection becomes exact. 
\par
\begin{figure}
\centering
\includegraphics[scale=1.0]{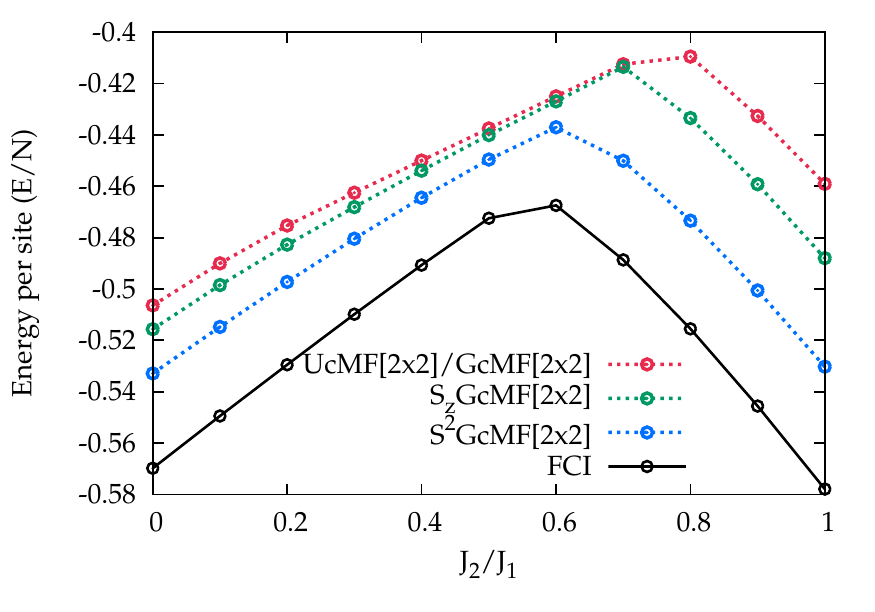}
\caption{Energy per site obtained in GcMF[$2 \times 2$], S$_z$GcMF[$2 \times 2$] and S$^2$GcMF[$2 \times 2$] for the quasi-1D $J_1 - J_2$ Heisenberg lattice.}
\label{s2_ladder}
\end{figure}

\begin{figure}
\centering
\includegraphics[scale=1.0]{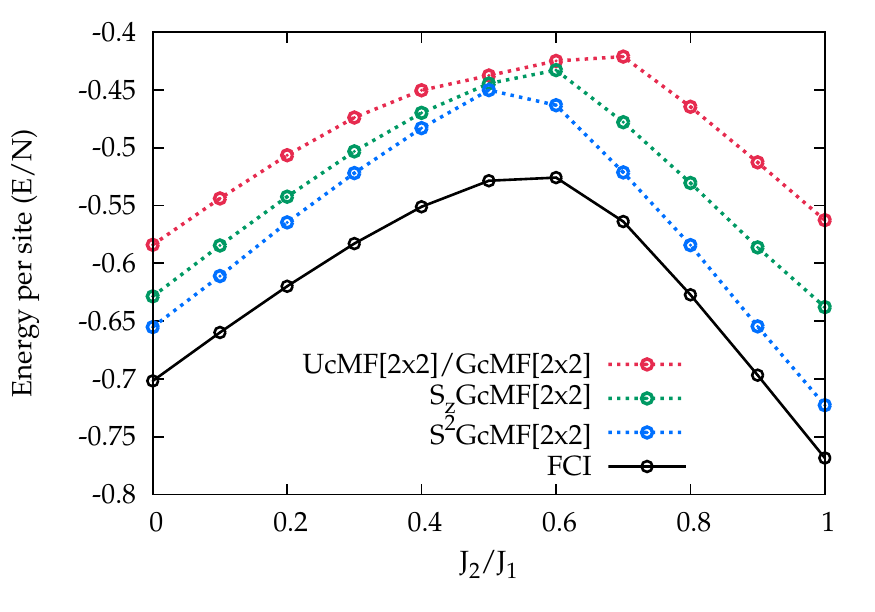}
\caption{Energy per site obtained in GcMF[$2 \times 2$], S$_z$GcMF[$2 \times 2$] and S$^2$GcMF[$2 \times 2$] for the 2D $J_1 - J_2$ Heisenberg lattice.}
\label{s2_square}
\end{figure}

 We conducted $S^2$ projection for the $2 \times 8$ and $4 \times 4$ $J_1 - J_2$ Heisenberg lattices, and our findings are summarized in Figs.~\ref{s2_ladder} and \ref{s2_square}, respectively. Overall, the $S^2$ projection effectively restores a significant portion of the missing correlations, particularly in the regimes of $J_2/J_1 < 0.4$ and $J_2/J_1 > 0.6$. This outcome aligns with expectations since the $S^2$ symmetry is considerably broken in those cases to capture the Néel and striped antiferromagnetic orders at the cMF. However, it is important to note that the computational cost associated with full spin projection is considerably higher and may not always be justified in certain scenarios, depending on the accuracy that needs to be achieved.

\section{Impact of cluster and system size on the energy}\label{app2}
\par To investigate the impact of cluster size and system size on the number of grid points required for accurate projection, we conducted two experiments; one for the one-dimensional $XXZ$ chain and one for the $J_1 - J_2$ ladder.
\par
\begin{figure}
\centering
\includegraphics[scale=1.0]{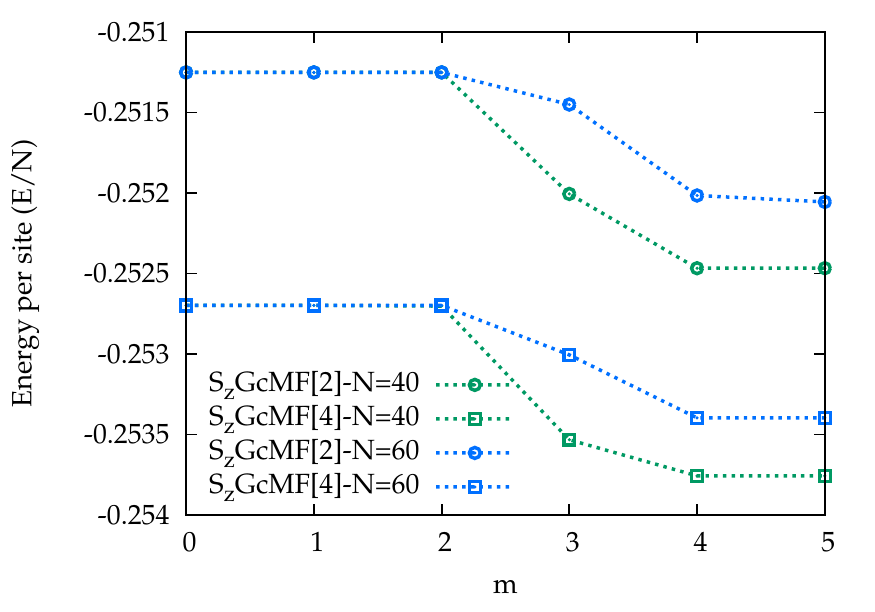}
\caption{S$_z$GcMF[2] energy per site in different $XXZ$ Hamiltonians as a function of the number of grid points used in the projection ($ n = 2^m$). The number after the dash refers to the size of the system.}
\label{cluslevels}
\end{figure}

\begin{figure}
\centering
\includegraphics[scale=1.0]{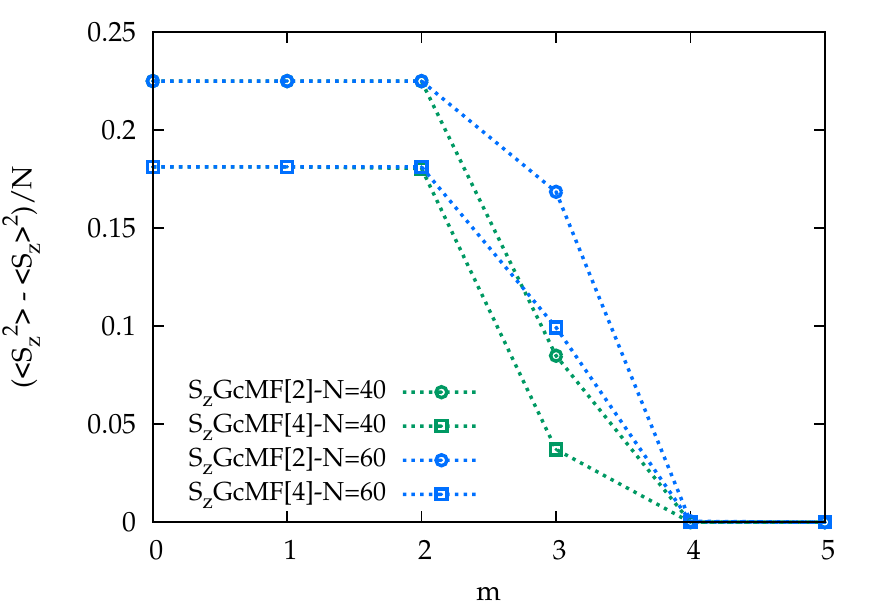}
\caption{Spin fluctuations obtained in S$_z$GcMF[2] for different system sizes of the $XXZ$ chain as a function of the number of grid points used in the projection ($ n = 2^m$). The number after the dash refers to the size of the system.}
\label{clusfluc}
\end{figure}

 For the first experiment, we specifically selected a value of the anisotropy parameter, $\Delta=-0.8$, where spin fluctuations are relatively strong. We performed two sets of calculations for two different system sizes, $N=40,60$ with periodic boundary conditions, in which we projected using grid points up to $\pi/16$. The first set used dimers (2-site clusters), while the second set used tetramers (4-site clusters). Our results are summarized in Figs.~\ref{cluslevels} and \ref{clusfluc}. We found that S$_z$cMF[4] required fewer grid points than S$_z$cMF[2], and this applied to both system sizes. This result was expected, as in the limit where the entire system is one cluster, symmetries are not broken, and therefore no grid points are required to restore them. Furthermore, we observed that larger systems required more grid points to reach full projection. This is due to the fact that for the same cluster size, the symmetry-broken state requires stronger spin fluctuations for a larger system to "exploit" its variational freedom to its limit. Consequently, more grid points are necessary to remove all these spin contaminants. Nevertheless, as we also observed for ladder systems in section \ref{4.3}, the significant spin contaminants and, therefore, the essential grid points, do not consistently coincide. As a result, the way by which the projection approaches the complete projection with respect to the number of grid points is not inherently evident or straightforward to determine.

\par
\begin{figure}
\centering
\includegraphics[scale=1.0]{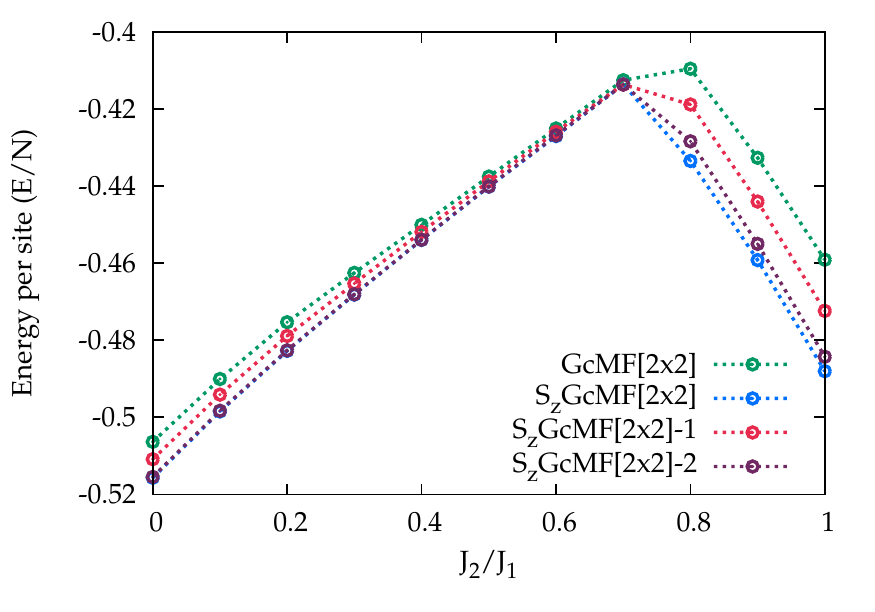}
\caption{Energy per site obtained in GcMF[$2 \times 2$] and \\ S$_z$GcMF[$2 \times 2$] for the $J_1 - J_2$ ladder as a function of the number of grid points used in the projection. The number after the dash refers to the number $m$ that corresponds to the number of grid points ($n = 2^m$) used for the projection (if there is no number, full projection is assumed).}
\label{levels_2}
\end{figure}

 For the second experiment, we aim to study how the energy gets impacted by the level of projection, for a system that does not break $S_z$ spontaneously. To explicate that, this time, we use a $2 \times 12$ $J_1 - J_2$ system, comprising six $2 \times 2$ clusters in a line with periodic boundary conditions (PBC). We present our general findings in Fig.~\ref{levels_2}. Our main observation is that the energy improvement is largest for the striped antiferromagnetic region ($J_2/J_1 \ge 0.8$). This phenomenon may occur because the striped spin configuration cannot be straightforwardly captured in $2 \times 2$ clusters, and therefore the NOCI part of the projection becomes more important. Overall, even for that system, when using finite systems, symmetry projection seems important. However, we need to remind the reader that for sufficiently large systems, the projection is energetically unimportant. 

\nocite{*}
\bibliography{aipsamp}

\providecommand{\noopsort}[1]{}\providecommand{\singleletter}[1]{#1}%
\begin{thebibliography}{104}%
\makeatletter
\providecommand \@ifxundefined [1]{%
 \@ifx{#1\undefined}
}%
\providecommand \@ifnum [1]{%
 \ifnum #1\expandafter \@firstoftwo
 \else \expandafter \@secondoftwo
 \fi
}%
\providecommand \@ifx [1]{%
 \ifx #1\expandafter \@firstoftwo
 \else \expandafter \@secondoftwo
 \fi
}%
\providecommand \natexlab [1]{#1}%
\providecommand \enquote  [1]{``#1''}%
\providecommand \bibnamefont  [1]{#1}%
\providecommand \bibfnamefont [1]{#1}%
\providecommand \citenamefont [1]{#1}%
\providecommand \href@noop [0]{\@secondoftwo}%
\providecommand \href [0]{\begingroup \@sanitize@url \@href}%
\providecommand \@href[1]{\@@startlink{#1}\@@href}%
\providecommand \@@href[1]{\endgroup#1\@@endlink}%
\providecommand \@sanitize@url [0]{\catcode `\\12\catcode `\$12\catcode
  `\&12\catcode `\#12\catcode `\^12\catcode `\_12\catcode `\%12\relax}%
\providecommand \@@startlink[1]{}%
\providecommand \@@endlink[0]{}%
\providecommand \url  [0]{\begingroup\@sanitize@url \@url }%
\providecommand \@url [1]{\endgroup\@href {#1}{\urlprefix }}%
\providecommand \urlprefix  [0]{URL }%
\providecommand \Eprint [0]{\href }%
\providecommand \doibase [0]{http://dx.doi.org/}%
\providecommand \selectlanguage [0]{\@gobble}%
\providecommand \bibinfo  [0]{\@secondoftwo}%
\providecommand \bibfield  [0]{\@secondoftwo}%
\providecommand \translation [1]{[#1]}%
\providecommand \BibitemOpen [0]{}%
\providecommand \bibitemStop [0]{}%
\providecommand \bibitemNoStop [0]{.\EOS\space}%
\providecommand \EOS [0]{\spacefactor3000\relax}%
\providecommand \BibitemShut  [1]{\csname bibitem#1\endcsname}%
\let\auto@bib@innerbib\@empty
\bibitem [{\citenamefont {Papastathopoulos-Katsaros}\ \emph
  {et~al.}(2022)\citenamefont {Papastathopoulos-Katsaros}, \citenamefont
  {Jiménez-Hoyos}, \citenamefont {Henderson},\ and\ \citenamefont
  {Scuseria}}]{papastathopoulos-katsaros_coupled_2022}%
  \BibitemOpen
  \bibfield  {author} {\bibinfo {author} {\bibfnamefont {A.}~\bibnamefont
  {Papastathopoulos-Katsaros}}, \bibinfo {author} {\bibfnamefont {C.~A.}\
  \bibnamefont {Jiménez-Hoyos}}, \bibinfo {author} {\bibfnamefont {T.~M.}\
  \bibnamefont {Henderson}}, \ and\ \bibinfo {author} {\bibfnamefont {G.~E.}\
  \bibnamefont {Scuseria}},\ }\bibfield  {title} {\enquote {\bibinfo {title}
  {Coupled {Cluster} and {Perturbation} {Theories} {Based} on a {Cluster}
  {Mean}-{Field} {Reference} {Applied} to {Strongly} {Correlated} {Spin}
  {Systems}},}\ }\href {\doibase 10.1021/acs.jctc.2c00338} {\bibfield
  {journal} {\bibinfo  {journal} {J. Chem. Theory Comput.}\ }\textbf {\bibinfo
  {volume} {18}},\ \bibinfo {pages} {4293--4303} (\bibinfo {year}
  {2022})}\BibitemShut {NoStop}%
\bibitem [{\citenamefont {Jiménez-Hoyos}\ and\ \citenamefont
  {Scuseria}(2015)}]{jimenez-hoyos_cluster-based_2015}%
  \BibitemOpen
  \bibfield  {author} {\bibinfo {author} {\bibfnamefont {C.~A.}\ \bibnamefont
  {Jiménez-Hoyos}}\ and\ \bibinfo {author} {\bibfnamefont {G.~E.}\
  \bibnamefont {Scuseria}},\ }\bibfield  {title} {\enquote {\bibinfo {title}
  {Cluster-based mean-field and perturbative description of strongly correlated
  fermion systems: {Application} to the one- and two-dimensional {Hubbard}
  model},}\ }\href {\doibase 10.1103/PhysRevB.92.085101} {\bibfield  {journal}
  {\bibinfo  {journal} {Phys. Rev. B}\ }\textbf {\bibinfo {volume} {92}},\
  \bibinfo {pages} {085101} (\bibinfo {year} {2015})}\BibitemShut {NoStop}%
\bibitem [{\citenamefont {Lykos}\ and\ \citenamefont
  {Pratt}(1963)}]{lykos_discussion_1963}%
  \BibitemOpen
  \bibfield  {author} {\bibinfo {author} {\bibfnamefont {P.}~\bibnamefont
  {Lykos}}\ and\ \bibinfo {author} {\bibfnamefont {G.~W.}\ \bibnamefont
  {Pratt}},\ }\bibfield  {title} {\enquote {\bibinfo {title} {Discussion on
  {The} {Hartree}-{Fock} {Approximation}},}\ }\href {\doibase
  10.1103/RevModPhys.35.496} {\bibfield  {journal} {\bibinfo  {journal} {Rev.
  Mod. Phys.}\ }\textbf {\bibinfo {volume} {35}},\ \bibinfo {pages} {496--501}
  (\bibinfo {year} {1963})}\BibitemShut {NoStop}%
\bibitem [{\citenamefont {Löwdin}\ and\ \citenamefont
  {Mayer}(1992)}]{lowdin_studies_1992}%
  \BibitemOpen
  \bibfield  {author} {\bibinfo {author} {\bibfnamefont {P.-O.}\ \bibnamefont
  {Löwdin}}\ and\ \bibinfo {author} {\bibfnamefont {I.}~\bibnamefont
  {Mayer}},\ }\bibfield  {title} {\enquote {\bibinfo {title} {Some {Studies} of
  the {General} {Hartree}-{Fock} {Method}},}\ }in\ \href {\doibase
  10.1016/S0065-3276(08)60101-X} {\emph {\bibinfo {booktitle} {Adv. Quantum
  Chem.}}},\ Vol.~\bibinfo {volume} {24}\ (\bibinfo  {publisher} {Elsevier},\
  \bibinfo {year} {1992})\ pp.\ \bibinfo {pages} {79--114}\BibitemShut
  {NoStop}%
\bibitem [{\citenamefont {Löwdin}(2007)}]{lefebvre_aspects_2007}%
  \BibitemOpen
  \bibfield  {author} {\bibinfo {author} {\bibfnamefont {P.-O.}\ \bibnamefont
  {Löwdin}},\ }\bibfield  {title} {\enquote {\bibinfo {title} {Some {Aspects}
  on the {Correlation} {Problem} and {Possible} {Extensions} of the
  {Independent}-{Particle} {Model}},}\ }in\ \href {\doibase
  10.1002/9780470143599.ch9} {\emph {\bibinfo {booktitle} {Adv. Chem.
  Phys.}}},\ \bibinfo {editor} {edited by\ \bibinfo {editor} {\bibfnamefont
  {R.}~\bibnamefont {LeFebvre}}\ and\ \bibinfo {editor} {\bibfnamefont
  {C.}~\bibnamefont {Moser}}}\ (\bibinfo  {publisher} {John Wiley \& Sons,
  Inc.},\ \bibinfo {address} {Hoboken, NJ, USA},\ \bibinfo {year} {2007})\ pp.\
  \bibinfo {pages} {283--340}\BibitemShut {NoStop}%
\bibitem [{\citenamefont {Valatin}(1961)}]{valatin_generalized_1961}%
  \BibitemOpen
  \bibfield  {author} {\bibinfo {author} {\bibfnamefont {J.~G.}\ \bibnamefont
  {Valatin}},\ }\bibfield  {title} {\enquote {\bibinfo {title} {Generalized
  {Hartree}-{Fock} {Method}},}\ }\href {\doibase 10.1103/PhysRev.122.1012}
  {\bibfield  {journal} {\bibinfo  {journal} {Phys. Rev.}\ }\textbf {\bibinfo
  {volume} {122}},\ \bibinfo {pages} {1012--1020} (\bibinfo {year}
  {1961})}\BibitemShut {NoStop}%
\bibitem [{\citenamefont {Jiménez-Hoyos}, \citenamefont {Henderson},\ and\
  \citenamefont {Scuseria}(2011)}]{jimenez-hoyos_generalized_2011}%
  \BibitemOpen
  \bibfield  {author} {\bibinfo {author} {\bibfnamefont {C.~A.}\ \bibnamefont
  {Jiménez-Hoyos}}, \bibinfo {author} {\bibfnamefont {T.~M.}\ \bibnamefont
  {Henderson}}, \ and\ \bibinfo {author} {\bibfnamefont {G.~E.}\ \bibnamefont
  {Scuseria}},\ }\bibfield  {title} {\enquote {\bibinfo {title} {Generalized
  {Hartree}–{Fock} {Description} of {Molecular} {Dissociation}},}\ }\href
  {\doibase 10.1021/ct200345a} {\bibfield  {journal} {\bibinfo  {journal} {J.
  Chem. Theory Comput.}\ }\textbf {\bibinfo {volume} {7}},\ \bibinfo {pages}
  {2667--2674} (\bibinfo {year} {2011})}\BibitemShut {NoStop}%
\bibitem [{\citenamefont {Hammes‐Schiffer}\ and\ \citenamefont
  {Andersen}(1993)}]{hammesschiffer_advantages_1993}%
  \BibitemOpen
  \bibfield  {author} {\bibinfo {author} {\bibfnamefont {S.}~\bibnamefont
  {Hammes‐Schiffer}}\ and\ \bibinfo {author} {\bibfnamefont {H.~C.}\
  \bibnamefont {Andersen}},\ }\bibfield  {title} {\enquote {\bibinfo {title}
  {The advantages of the general {Hartree}–{Fock} method for future computer
  simulation of materials},}\ }\href {\doibase 10.1063/1.465305} {\bibfield
  {journal} {\bibinfo  {journal} {J. Chem. Phys.}\ }\textbf {\bibinfo {volume}
  {99}},\ \bibinfo {pages} {1901--1913} (\bibinfo {year} {1993})}\BibitemShut
  {NoStop}%
\bibitem [{\citenamefont {Jake}, \citenamefont {Henderson},\ and\ \citenamefont
  {Scuseria}(2018)}]{jake_hartreefock_2018}%
  \BibitemOpen
  \bibfield  {author} {\bibinfo {author} {\bibfnamefont {L.~C.}\ \bibnamefont
  {Jake}}, \bibinfo {author} {\bibfnamefont {T.~M.}\ \bibnamefont {Henderson}},
  \ and\ \bibinfo {author} {\bibfnamefont {G.~E.}\ \bibnamefont {Scuseria}},\
  }\bibfield  {title} {\enquote {\bibinfo {title} {Hartree–{Fock} symmetry
  breaking around conical intersections},}\ }\href {\doibase 10.1063/1.5010929}
  {\bibfield  {journal} {\bibinfo  {journal} {J. Chem. Phys.}\ }\textbf
  {\bibinfo {volume} {148}},\ \bibinfo {pages} {024109} (\bibinfo {year}
  {2018})}\BibitemShut {NoStop}%
\bibitem [{\citenamefont {Ring}\ and\ \citenamefont
  {Schuck}(2004)}]{ring_nuclear_2004}%
  \BibitemOpen
  \bibfield  {author} {\bibinfo {author} {\bibfnamefont {P.}~\bibnamefont
  {Ring}}\ and\ \bibinfo {author} {\bibfnamefont {P.}~\bibnamefont {Schuck}},\
  }\href@noop {} {\emph {\bibinfo {title} {The nuclear many body problem}}}\
  (\bibinfo  {publisher} {Springer Science \& Business Media},\ \bibinfo {year}
  {2004})\BibitemShut {NoStop}%
\bibitem [{\citenamefont {Blaizot}\ and\ \citenamefont
  {Ripka}(1986)}]{blaizot_quantum_1986}%
  \BibitemOpen
  \bibfield  {author} {\bibinfo {author} {\bibfnamefont {J.-P.}\ \bibnamefont
  {Blaizot}}\ and\ \bibinfo {author} {\bibfnamefont {G.}~\bibnamefont
  {Ripka}},\ }\href@noop {} {\emph {\bibinfo {title} {Quantum theory of finite
  systems}}}\ (\bibinfo  {publisher} {MIT Press},\ \bibinfo {address}
  {Cambridge, MA},\ \bibinfo {year} {1986})\BibitemShut {NoStop}%
\bibitem [{\citenamefont {Löwdin}(1955)}]{lowdin_quantum_1955}%
  \BibitemOpen
  \bibfield  {author} {\bibinfo {author} {\bibfnamefont {P.-O.}\ \bibnamefont
  {Löwdin}},\ }\bibfield  {title} {\enquote {\bibinfo {title} {Quantum
  {Theory} of {Many}-{Particle} {Systems}. {III}. {Extension} of the
  {Hartree}-{Fock} {Scheme} to {Include} {Degenerate} {Systems} and
  {Correlation} {Effects}},}\ }\href {\doibase 10.1103/PhysRev.97.1509}
  {\bibfield  {journal} {\bibinfo  {journal} {Phys. Rev.}\ }\textbf {\bibinfo
  {volume} {97}},\ \bibinfo {pages} {1509--1520} (\bibinfo {year}
  {1955})}\BibitemShut {NoStop}%
\bibitem [{\citenamefont {Schmid}(2004)}]{schmid_use_2004}%
  \BibitemOpen
  \bibfield  {author} {\bibinfo {author} {\bibfnamefont {K.~W.}\ \bibnamefont
  {Schmid}},\ }\bibfield  {title} {\enquote {\bibinfo {title} {On the use of
  general symmetry-projected {Hartree}–{Fock}–{Bogoliubov} configurations
  in variational approaches to the nuclear many-body problem},}\ }\href
  {\doibase 10.1016/j.ppnp.2004.02.001} {\bibfield  {journal} {\bibinfo
  {journal} {Prog. Part. Nucl. Phys.}\ }\textbf {\bibinfo {volume} {52}},\
  \bibinfo {pages} {565--633} (\bibinfo {year} {2004})}\BibitemShut {NoStop}%
\bibitem [{\citenamefont {Jiménez-Hoyos}\ \emph {et~al.}(2012)\citenamefont
  {Jiménez-Hoyos}, \citenamefont {Henderson}, \citenamefont {Tsuchimochi},\
  and\ \citenamefont {Scuseria}}]{jimenez-hoyos_projected_2012}%
  \BibitemOpen
  \bibfield  {author} {\bibinfo {author} {\bibfnamefont {C.~A.}\ \bibnamefont
  {Jiménez-Hoyos}}, \bibinfo {author} {\bibfnamefont {T.~M.}\ \bibnamefont
  {Henderson}}, \bibinfo {author} {\bibfnamefont {T.}~\bibnamefont
  {Tsuchimochi}}, \ and\ \bibinfo {author} {\bibfnamefont {G.~E.}\ \bibnamefont
  {Scuseria}},\ }\bibfield  {title} {\enquote {\bibinfo {title} {Projected
  {Hartree}–{Fock} theory},}\ }\href {\doibase 10.1063/1.4705280} {\bibfield
  {journal} {\bibinfo  {journal} {J. Chem. Phys.}\ }\textbf {\bibinfo {volume}
  {136}},\ \bibinfo {pages} {164109} (\bibinfo {year} {2012})}\BibitemShut
  {NoStop}%
\bibitem [{\citenamefont {Ghassemi~Tabrizi}\ and\ \citenamefont
  {Jiménez-Hoyos}(2022)}]{ghassemi_tabrizi_ground_2022}%
  \BibitemOpen
  \bibfield  {author} {\bibinfo {author} {\bibfnamefont {S.}~\bibnamefont
  {Ghassemi~Tabrizi}}\ and\ \bibinfo {author} {\bibfnamefont {C.~A.}\
  \bibnamefont {Jiménez-Hoyos}},\ }\bibfield  {title} {\enquote {\bibinfo
  {title} {Ground states of {Heisenberg} spin clusters from projected
  {Hartree}-{Fock} theory},}\ }\href {\doibase 10.1103/PhysRevB.105.035147}
  {\bibfield  {journal} {\bibinfo  {journal} {Phys. Rev. B}\ }\textbf {\bibinfo
  {volume} {105}},\ \bibinfo {pages} {035147} (\bibinfo {year}
  {2022})}\BibitemShut {NoStop}%
\bibitem [{\citenamefont {Ruiz}(2022)}]{ruiz_halfprojected_2022}%
  \BibitemOpen
  \bibfield  {author} {\bibinfo {author} {\bibfnamefont {M.~B.}\ \bibnamefont
  {Ruiz}},\ }\bibfield  {title} {\enquote {\bibinfo {title} {Half‐{Projected}
  {Hartree}–{Fock} method: {History} and application to excited states of the
  same symmetry as the ground state},}\ }\href
  {https://onlinelibrary.wiley.com/doi/10.1002/qua.26889} {\bibfield  {journal}
  {\bibinfo  {journal} {Int. J. Quantum Chem.}\ }\textbf {\bibinfo {volume}
  {122}} (\bibinfo {year} {2022})}\BibitemShut {NoStop}%
\bibitem [{\citenamefont {Henderson}\ and\ \citenamefont
  {Scuseria}(2017)}]{henderson_spin-projected_2017}%
  \BibitemOpen
  \bibfield  {author} {\bibinfo {author} {\bibfnamefont {T.~M.}\ \bibnamefont
  {Henderson}}\ and\ \bibinfo {author} {\bibfnamefont {G.~E.}\ \bibnamefont
  {Scuseria}},\ }\bibfield  {title} {\enquote {\bibinfo {title} {Spin-projected
  generalized {Hartree}-{Fock} method as a polynomial of particle-hole
  excitations},}\ }\href {\doibase 10.1103/PhysRevA.96.022506} {\bibfield
  {journal} {\bibinfo  {journal} {Phys. Rev. A}\ }\textbf {\bibinfo {volume}
  {96}},\ \bibinfo {pages} {022506} (\bibinfo {year} {2017})}\BibitemShut
  {NoStop}%
\bibitem [{\citenamefont {Garza}, \citenamefont {Jiménez-Hoyos},\ and\
  \citenamefont {Scuseria}(2014)}]{garza_electronic_2014}%
  \BibitemOpen
  \bibfield  {author} {\bibinfo {author} {\bibfnamefont {A.~J.}\ \bibnamefont
  {Garza}}, \bibinfo {author} {\bibfnamefont {C.~A.}\ \bibnamefont
  {Jiménez-Hoyos}}, \ and\ \bibinfo {author} {\bibfnamefont {G.~E.}\
  \bibnamefont {Scuseria}},\ }\bibfield  {title} {\enquote {\bibinfo {title}
  {Electronic correlation without double counting via a combination of spin
  projected {Hartree}-{Fock} and density functional theories},}\ }\href
  {\doibase 10.1063/1.4883491} {\bibfield  {journal} {\bibinfo  {journal} {J.
  Chem. Phys.}\ }\textbf {\bibinfo {volume} {140}},\ \bibinfo {pages} {244102}
  (\bibinfo {year} {2014})}\BibitemShut {NoStop}%
\bibitem [{\citenamefont {Shi}\ \emph {et~al.}(2014)\citenamefont {Shi},
  \citenamefont {Jiménez-Hoyos}, \citenamefont {Rodríguez-Guzmán},
  \citenamefont {Scuseria},\ and\ \citenamefont
  {Zhang}}]{shi_symmetry-projected_2014}%
  \BibitemOpen
  \bibfield  {author} {\bibinfo {author} {\bibfnamefont {H.}~\bibnamefont
  {Shi}}, \bibinfo {author} {\bibfnamefont {C.~A.}\ \bibnamefont
  {Jiménez-Hoyos}}, \bibinfo {author} {\bibfnamefont {R.}~\bibnamefont
  {Rodríguez-Guzmán}}, \bibinfo {author} {\bibfnamefont {G.~E.}\ \bibnamefont
  {Scuseria}}, \ and\ \bibinfo {author} {\bibfnamefont {S.}~\bibnamefont
  {Zhang}},\ }\bibfield  {title} {\enquote {\bibinfo {title}
  {Symmetry-projected wave functions in quantum {Monte} {Carlo}
  calculations},}\ }\href {\doibase 10.1103/PhysRevB.89.125129} {\bibfield
  {journal} {\bibinfo  {journal} {Phys. Rev. B}\ }\textbf {\bibinfo {volume}
  {89}},\ \bibinfo {pages} {125129} (\bibinfo {year} {2014})}\BibitemShut
  {NoStop}%
\bibitem [{\citenamefont {Cui}\ \emph {et~al.}(2013)\citenamefont {Cui},
  \citenamefont {Bulik}, \citenamefont {Jiménez-Hoyos}, \citenamefont
  {Henderson},\ and\ \citenamefont {Scuseria}}]{cui_proper_2013}%
  \BibitemOpen
  \bibfield  {author} {\bibinfo {author} {\bibfnamefont {Y.}~\bibnamefont
  {Cui}}, \bibinfo {author} {\bibfnamefont {I.~W.}\ \bibnamefont {Bulik}},
  \bibinfo {author} {\bibfnamefont {C.~A.}\ \bibnamefont {Jiménez-Hoyos}},
  \bibinfo {author} {\bibfnamefont {T.~M.}\ \bibnamefont {Henderson}}, \ and\
  \bibinfo {author} {\bibfnamefont {G.~E.}\ \bibnamefont {Scuseria}},\
  }\bibfield  {title} {\enquote {\bibinfo {title} {Proper and improper zero
  energy modes in {Hartree}-{Fock} theory and their relevance for symmetry
  breaking and restoration},}\ }\href {\doibase 10.1063/1.4824905} {\bibfield
  {journal} {\bibinfo  {journal} {J. Chem. Phys.}\ }\textbf {\bibinfo {volume}
  {139}},\ \bibinfo {pages} {154107} (\bibinfo {year} {2013})}\BibitemShut
  {NoStop}%
\bibitem [{\citenamefont {Tsuchimochi}\ and\ \citenamefont
  {Ten-no}(2017)}]{tsuchimochi_bridging_2017}%
  \BibitemOpen
  \bibfield  {author} {\bibinfo {author} {\bibfnamefont {T.}~\bibnamefont
  {Tsuchimochi}}\ and\ \bibinfo {author} {\bibfnamefont {S.}~\bibnamefont
  {Ten-no}},\ }\bibfield  {title} {\enquote {\bibinfo {title} {Bridging
  {Single}- and {Multireference} {Domains} for {Electron} {Correlation}:
  {Spin}-{Extended} {Coupled} {Electron} {Pair} {Approximation}},}\ }\href
  {\doibase 10.1021/acs.jctc.7b00073} {\bibfield  {journal} {\bibinfo
  {journal} {J. Chem. Theory Comput.}\ }\textbf {\bibinfo {volume} {13}},\
  \bibinfo {pages} {1667--1681} (\bibinfo {year} {2017})}\BibitemShut {NoStop}%
\bibitem [{\citenamefont {Tsuchimochi}\ and\ \citenamefont
  {Ten-no}(2016{\natexlab{a}})}]{tsuchimochi_black-box_2016}%
  \BibitemOpen
  \bibfield  {author} {\bibinfo {author} {\bibfnamefont {T.}~\bibnamefont
  {Tsuchimochi}}\ and\ \bibinfo {author} {\bibfnamefont {S.}~\bibnamefont
  {Ten-no}},\ }\bibfield  {title} {\enquote {\bibinfo {title} {Black-{Box}
  {Description} of {Electron} {Correlation} with the {Spin}-{Extended}
  {Configuration} {Interaction} {Model}: {Implementation} and {Assessment}},}\
  }\href {\doibase 10.1021/acs.jctc.6b00137} {\bibfield  {journal} {\bibinfo
  {journal} {J. Chem. Theory Comput.}\ }\textbf {\bibinfo {volume} {12}},\
  \bibinfo {pages} {1741--1759} (\bibinfo {year}
  {2016}{\natexlab{a}})}\BibitemShut {NoStop}%
\bibitem [{\citenamefont {Tsuchimochi}\ and\ \citenamefont
  {Ten-no}(2016{\natexlab{b}})}]{tsuchimochi_communication_2016}%
  \BibitemOpen
  \bibfield  {author} {\bibinfo {author} {\bibfnamefont {T.}~\bibnamefont
  {Tsuchimochi}}\ and\ \bibinfo {author} {\bibfnamefont {S.}~\bibnamefont
  {Ten-no}},\ }\bibfield  {title} {\enquote {\bibinfo {title} {Communication:
  {Configuration} interaction combined with spin-projection for strongly
  correlated molecular electronic structures},}\ }\href {\doibase
  10.1063/1.4939585} {\bibfield  {journal} {\bibinfo  {journal} {J. Chem.
  Phys.}\ }\textbf {\bibinfo {volume} {144}},\ \bibinfo {pages} {011101}
  (\bibinfo {year} {2016}{\natexlab{b}})}\BibitemShut {NoStop}%
\bibitem [{\citenamefont {Schlegel}(1988)}]{schlegel_moeller-plesset_1988}%
  \BibitemOpen
  \bibfield  {author} {\bibinfo {author} {\bibfnamefont {H.~B.}\ \bibnamefont
  {Schlegel}},\ }\bibfield  {title} {\enquote {\bibinfo {title}
  {{M}øller-{Plesset} perturbation theory with spin projection},}\ }\href
  {\doibase 10.1021/j100322a014} {\bibfield  {journal} {\bibinfo  {journal} {J.
  Phys. Chem.}\ }\textbf {\bibinfo {volume} {92}},\ \bibinfo {pages}
  {3075--3078} (\bibinfo {year} {1988})}\BibitemShut {NoStop}%
\bibitem [{\citenamefont {Schlegel}(1986)}]{schlegel_potential_1986}%
  \BibitemOpen
  \bibfield  {author} {\bibinfo {author} {\bibfnamefont {H.~B.}\ \bibnamefont
  {Schlegel}},\ }\bibfield  {title} {\enquote {\bibinfo {title} {Potential
  energy curves using unrestricted {M}øller–{Plesset} perturbation theory
  with spin annihilation},}\ }\href {\doibase 10.1063/1.450026} {\bibfield
  {journal} {\bibinfo  {journal} {J. Chem. Phys.}\ }\textbf {\bibinfo {volume}
  {84}},\ \bibinfo {pages} {4530--4534} (\bibinfo {year} {1986})}\BibitemShut
  {NoStop}%
\bibitem [{\citenamefont {Knowles}\ and\ \citenamefont
  {Handy}(1988{\natexlab{a}})}]{knowles_convergence_1988}%
  \BibitemOpen
  \bibfield  {author} {\bibinfo {author} {\bibfnamefont {P.~J.}\ \bibnamefont
  {Knowles}}\ and\ \bibinfo {author} {\bibfnamefont {N.~C.}\ \bibnamefont
  {Handy}},\ }\bibfield  {title} {\enquote {\bibinfo {title} {Convergence of
  projected unrestricted {Hartee}-{Fock} {M}øller-{Plesset} series.}}\ }\href
  {\doibase 10.1021/j100322a018} {\bibfield  {journal} {\bibinfo  {journal} {J.
  Phys. Chem.}\ }\textbf {\bibinfo {volume} {92}},\ \bibinfo {pages}
  {3097--3100} (\bibinfo {year} {1988}{\natexlab{a}})}\BibitemShut {NoStop}%
\bibitem [{\citenamefont {Knowles}\ and\ \citenamefont
  {Handy}(1988{\natexlab{b}})}]{knowles_projected_1988}%
  \BibitemOpen
  \bibfield  {author} {\bibinfo {author} {\bibfnamefont {P.~J.}\ \bibnamefont
  {Knowles}}\ and\ \bibinfo {author} {\bibfnamefont {N.~C.}\ \bibnamefont
  {Handy}},\ }\bibfield  {title} {\enquote {\bibinfo {title} {Projected
  unrestricted {M}øller–{Plesset} second‐order energies},}\ }\href
  {\doibase 10.1063/1.454397} {\bibfield  {journal} {\bibinfo  {journal} {J.
  Chem. Phys.}\ }\textbf {\bibinfo {volume} {88}},\ \bibinfo {pages}
  {6991--6998} (\bibinfo {year} {1988}{\natexlab{b}})}\BibitemShut {NoStop}%
\bibitem [{\citenamefont {Qiu}\ \emph {et~al.}(2017)\citenamefont {Qiu},
  \citenamefont {Henderson}, \citenamefont {Zhao},\ and\ \citenamefont
  {Scuseria}}]{qiuu_projected_2017}%
  \BibitemOpen
  \bibfield  {author} {\bibinfo {author} {\bibfnamefont {Y.}~\bibnamefont
  {Qiu}}, \bibinfo {author} {\bibfnamefont {T.~M.}\ \bibnamefont {Henderson}},
  \bibinfo {author} {\bibfnamefont {J.}~\bibnamefont {Zhao}}, \ and\ \bibinfo
  {author} {\bibfnamefont {G.~E.}\ \bibnamefont {Scuseria}},\ }\bibfield
  {title} {\enquote {\bibinfo {title} {Projected coupled cluster theory},}\
  }\href {\doibase 10.1063/1.4991020} {\bibfield  {journal} {\bibinfo
  {journal} {J. Chem. Phys.}\ }\textbf {\bibinfo {volume} {147}},\ \bibinfo
  {pages} {064111} (\bibinfo {year} {2017})}\BibitemShut {NoStop}%
\bibitem [{\citenamefont {Qiu}, \citenamefont {Henderson},\ and\ \citenamefont
  {Scuseria}(2017)}]{qiu_projected_2017}%
  \BibitemOpen
  \bibfield  {author} {\bibinfo {author} {\bibfnamefont {Y.}~\bibnamefont
  {Qiu}}, \bibinfo {author} {\bibfnamefont {T.~M.}\ \bibnamefont {Henderson}},
  \ and\ \bibinfo {author} {\bibfnamefont {G.~E.}\ \bibnamefont {Scuseria}},\
  }\bibfield  {title} {\enquote {\bibinfo {title} {Projected {Hartree}-{Fock}
  theory as a polynomial of particle-hole excitations and its combination with
  variational coupled cluster theory},}\ }\href {\doibase 10.1063/1.4983065}
  {\bibfield  {journal} {\bibinfo  {journal} {J. Chem. Phys.}\ }\textbf
  {\bibinfo {volume} {146}},\ \bibinfo {pages} {184105} (\bibinfo {year}
  {2017})}\BibitemShut {NoStop}%
\bibitem [{\citenamefont {Gomez}, \citenamefont {Henderson},\ and\
  \citenamefont {Scuseria}(2017)}]{gomez_attenuated_2017}%
  \BibitemOpen
  \bibfield  {author} {\bibinfo {author} {\bibfnamefont {J.~A.}\ \bibnamefont
  {Gomez}}, \bibinfo {author} {\bibfnamefont {T.~M.}\ \bibnamefont
  {Henderson}}, \ and\ \bibinfo {author} {\bibfnamefont {G.~E.}\ \bibnamefont
  {Scuseria}},\ }\bibfield  {title} {\enquote {\bibinfo {title} {Attenuated
  coupled cluster: a heuristic polynomial similarity transformation
  incorporating spin symmetry projection into traditional coupled cluster
  theory},}\ }\href {\doibase 10.1080/00268976.2017.1302610} {\bibfield
  {journal} {\bibinfo  {journal} {Mol. Phys.}\ }\textbf {\bibinfo {volume}
  {115}},\ \bibinfo {pages} {2673--2683} (\bibinfo {year} {2017})}\BibitemShut
  {NoStop}%
\bibitem [{\citenamefont {Qiu}\ \emph {et~al.}(2018)\citenamefont {Qiu},
  \citenamefont {Henderson}, \citenamefont {Zhao},\ and\ \citenamefont
  {Scuseria}}]{qiu_projected_2018}%
  \BibitemOpen
  \bibfield  {author} {\bibinfo {author} {\bibfnamefont {Y.}~\bibnamefont
  {Qiu}}, \bibinfo {author} {\bibfnamefont {T.~M.}\ \bibnamefont {Henderson}},
  \bibinfo {author} {\bibfnamefont {J.}~\bibnamefont {Zhao}}, \ and\ \bibinfo
  {author} {\bibfnamefont {G.~E.}\ \bibnamefont {Scuseria}},\ }\bibfield
  {title} {\enquote {\bibinfo {title} {Projected coupled cluster theory:
  {Optimization} of cluster amplitudes in the presence of symmetry
  projection},}\ }\href {\doibase 10.1063/1.5053605} {\bibfield  {journal}
  {\bibinfo  {journal} {J. Chem. Phys.}\ }\textbf {\bibinfo {volume} {149}},\
  \bibinfo {pages} {164108} (\bibinfo {year} {2018})}\BibitemShut {NoStop}%
\bibitem [{\citenamefont {Gomez}, \citenamefont {Henderson},\ and\
  \citenamefont {Scuseria}(2019)}]{gomez_polynomial-product_2019}%
  \BibitemOpen
  \bibfield  {author} {\bibinfo {author} {\bibfnamefont {J.~A.}\ \bibnamefont
  {Gomez}}, \bibinfo {author} {\bibfnamefont {T.~M.}\ \bibnamefont
  {Henderson}}, \ and\ \bibinfo {author} {\bibfnamefont {G.~E.}\ \bibnamefont
  {Scuseria}},\ }\bibfield  {title} {\enquote {\bibinfo {title}
  {Polynomial-product states: {A} symmetry-projection-based factorization of
  the full coupled cluster wavefunction in terms of polynomials of double
  excitations},}\ }\href {\doibase 10.1063/1.5085314} {\bibfield  {journal}
  {\bibinfo  {journal} {J. Chem. Phys.}\ }\textbf {\bibinfo {volume} {150}},\
  \bibinfo {pages} {144108} (\bibinfo {year} {2019})}\BibitemShut {NoStop}%
\bibitem [{\citenamefont {Wahlen-Strothman}\ \emph {et~al.}(2017)\citenamefont
  {Wahlen-Strothman}, \citenamefont {Henderson}, \citenamefont {Hermes},
  \citenamefont {Degroote}, \citenamefont {Qiu}, \citenamefont {Zhao},
  \citenamefont {Dukelsky},\ and\ \citenamefont
  {Scuseria}}]{wahlen-strothman_merging_2017}%
  \BibitemOpen
  \bibfield  {author} {\bibinfo {author} {\bibfnamefont {J.~M.}\ \bibnamefont
  {Wahlen-Strothman}}, \bibinfo {author} {\bibfnamefont {T.~M.}\ \bibnamefont
  {Henderson}}, \bibinfo {author} {\bibfnamefont {M.~R.}\ \bibnamefont
  {Hermes}}, \bibinfo {author} {\bibfnamefont {M.}~\bibnamefont {Degroote}},
  \bibinfo {author} {\bibfnamefont {Y.}~\bibnamefont {Qiu}}, \bibinfo {author}
  {\bibfnamefont {J.}~\bibnamefont {Zhao}}, \bibinfo {author} {\bibfnamefont
  {J.}~\bibnamefont {Dukelsky}}, \ and\ \bibinfo {author} {\bibfnamefont
  {G.~E.}\ \bibnamefont {Scuseria}},\ }\bibfield  {title} {\enquote {\bibinfo
  {title} {Merging symmetry projection methods with coupled cluster theory:
  {Lessons} from the {Lipkin} model {Hamiltonian}},}\ }\href {\doibase
  10.1063/1.4974989} {\bibfield  {journal} {\bibinfo  {journal} {J. Chem.
  Phys.}\ }\textbf {\bibinfo {volume} {146}},\ \bibinfo {pages} {054110}
  (\bibinfo {year} {2017})}\BibitemShut {NoStop}%
\bibitem [{\citenamefont {Rodríguez-Guzmán}, \citenamefont {Jiménez-Hoyos},\
  and\ \citenamefont
  {Scuseria}(2014{\natexlab{a}})}]{rodriguez-guzman_variational_2014}%
  \BibitemOpen
  \bibfield  {author} {\bibinfo {author} {\bibfnamefont {R.}~\bibnamefont
  {Rodríguez-Guzmán}}, \bibinfo {author} {\bibfnamefont {C.~A.}\ \bibnamefont
  {Jiménez-Hoyos}}, \ and\ \bibinfo {author} {\bibfnamefont {G.~E.}\
  \bibnamefont {Scuseria}},\ }\bibfield  {title} {\enquote {\bibinfo {title}
  {Variational description of the ground state of the repulsive two-dimensional
  {Hubbard} model in terms of nonorthogonal symmetry-projected {Slater}
  determinants},}\ }\href {\doibase 10.1103/PhysRevB.90.195110} {\bibfield
  {journal} {\bibinfo  {journal} {Phys. Rev. B}\ }\textbf {\bibinfo {volume}
  {90}},\ \bibinfo {pages} {195110} (\bibinfo {year}
  {2014}{\natexlab{a}})}\BibitemShut {NoStop}%
\bibitem [{\citenamefont {Bytautas}\ \emph {et~al.}(2014)\citenamefont
  {Bytautas}, \citenamefont {Jiménez-Hoyos}, \citenamefont
  {Rodríguez-Guzmán},\ and\ \citenamefont
  {Scuseria}}]{bytautas_potential_2014}%
  \BibitemOpen
  \bibfield  {author} {\bibinfo {author} {\bibfnamefont {L.}~\bibnamefont
  {Bytautas}}, \bibinfo {author} {\bibfnamefont {C.~A.}\ \bibnamefont
  {Jiménez-Hoyos}}, \bibinfo {author} {\bibfnamefont {R.}~\bibnamefont
  {Rodríguez-Guzmán}}, \ and\ \bibinfo {author} {\bibfnamefont {G.~E.}\
  \bibnamefont {Scuseria}},\ }\bibfield  {title} {\enquote {\bibinfo {title}
  {Potential energy curves for {Mo}$_{\textrm{2}}$ : multi-component
  symmetry-projected {Hartree}–{Fock} and beyond},}\ }\href {\doibase
  10.1080/00268976.2013.874623} {\bibfield  {journal} {\bibinfo  {journal}
  {Mol. Phys.}\ }\textbf {\bibinfo {volume} {112}},\ \bibinfo {pages}
  {1938--1946} (\bibinfo {year} {2014})}\BibitemShut {NoStop}%
\bibitem [{\citenamefont {Rodríguez-Guzmán}\ \emph
  {et~al.}(2013)\citenamefont {Rodríguez-Guzmán}, \citenamefont
  {Jiménez-Hoyos}, \citenamefont {Schutski},\ and\ \citenamefont
  {Scuseria}}]{rodriguez-guzman_multireference_2013}%
  \BibitemOpen
  \bibfield  {author} {\bibinfo {author} {\bibfnamefont {R.}~\bibnamefont
  {Rodríguez-Guzmán}}, \bibinfo {author} {\bibfnamefont {C.~A.}\ \bibnamefont
  {Jiménez-Hoyos}}, \bibinfo {author} {\bibfnamefont {R.}~\bibnamefont
  {Schutski}}, \ and\ \bibinfo {author} {\bibfnamefont {G.~E.}\ \bibnamefont
  {Scuseria}},\ }\bibfield  {title} {\enquote {\bibinfo {title} {Multireference
  symmetry-projected variational approaches for ground and excited states of
  the one-dimensional {Hubbard} model},}\ }\href {\doibase
  10.1103/PhysRevB.87.235129} {\bibfield  {journal} {\bibinfo  {journal} {Phys.
  Rev. B}\ }\textbf {\bibinfo {volume} {87}},\ \bibinfo {pages} {235129}
  (\bibinfo {year} {2013})}\BibitemShut {NoStop}%
\bibitem [{\citenamefont {Jiménez-Hoyos}, \citenamefont {Rodríguez-Guzmán},\
  and\ \citenamefont {Scuseria}(2013)}]{jimenez-hoyos_multi-component_2013}%
  \BibitemOpen
  \bibfield  {author} {\bibinfo {author} {\bibfnamefont {C.~A.}\ \bibnamefont
  {Jiménez-Hoyos}}, \bibinfo {author} {\bibfnamefont {R.}~\bibnamefont
  {Rodríguez-Guzmán}}, \ and\ \bibinfo {author} {\bibfnamefont {G.~E.}\
  \bibnamefont {Scuseria}},\ }\bibfield  {title} {\enquote {\bibinfo {title}
  {Multi-component symmetry-projected approach for molecular ground state
  correlations},}\ }\href {\doibase 10.1063/1.4832476} {\bibfield  {journal}
  {\bibinfo  {journal} {J. Chem. Phys.}\ }\textbf {\bibinfo {volume} {139}},\
  \bibinfo {pages} {204102} (\bibinfo {year} {2013})}\BibitemShut {NoStop}%
\bibitem [{\citenamefont {Rodríguez-Guzmán}, \citenamefont {Jiménez-Hoyos},\
  and\ \citenamefont
  {Scuseria}(2014{\natexlab{b}})}]{rodriguez-guzman_multireference_2014}%
  \BibitemOpen
  \bibfield  {author} {\bibinfo {author} {\bibfnamefont {R.}~\bibnamefont
  {Rodríguez-Guzmán}}, \bibinfo {author} {\bibfnamefont {C.~A.}\ \bibnamefont
  {Jiménez-Hoyos}}, \ and\ \bibinfo {author} {\bibfnamefont {G.~E.}\
  \bibnamefont {Scuseria}},\ }\bibfield  {title} {\enquote {\bibinfo {title}
  {Multireference symmetry-projected variational approximation for the ground
  state of the doped one-dimensional {Hubbard} model},}\ }\href {\doibase
  10.1103/PhysRevB.89.195109} {\bibfield  {journal} {\bibinfo  {journal} {Phys.
  Rev. B}\ }\textbf {\bibinfo {volume} {89}},\ \bibinfo {pages} {195109}
  (\bibinfo {year} {2014}{\natexlab{b}})}\BibitemShut {NoStop}%
\bibitem [{\citenamefont {Rodríguez-Guzmán}\ \emph
  {et~al.}(2012)\citenamefont {Rodríguez-Guzmán}, \citenamefont {Schmid},
  \citenamefont {Jiménez-Hoyos},\ and\ \citenamefont
  {Scuseria}}]{rodriguez-guzman_symmetry-projected_2012}%
  \BibitemOpen
  \bibfield  {author} {\bibinfo {author} {\bibfnamefont {R.}~\bibnamefont
  {Rodríguez-Guzmán}}, \bibinfo {author} {\bibfnamefont {K.~W.}\ \bibnamefont
  {Schmid}}, \bibinfo {author} {\bibfnamefont {C.~A.}\ \bibnamefont
  {Jiménez-Hoyos}}, \ and\ \bibinfo {author} {\bibfnamefont {G.~E.}\
  \bibnamefont {Scuseria}},\ }\bibfield  {title} {\enquote {\bibinfo {title}
  {Symmetry-projected variational approach for ground and excited states of the
  two-dimensional {Hubbard} model},}\ }\href {\doibase
  10.1103/PhysRevB.85.245130} {\bibfield  {journal} {\bibinfo  {journal} {Phys.
  Rev. B}\ }\textbf {\bibinfo {volume} {85}},\ \bibinfo {pages} {245130}
  (\bibinfo {year} {2012})}\BibitemShut {NoStop}%
\bibitem [{\citenamefont {Li}(2004)}]{li_block-correlated_2004}%
  \BibitemOpen
  \bibfield  {author} {\bibinfo {author} {\bibfnamefont {S.}~\bibnamefont
  {Li}},\ }\bibfield  {title} {\enquote {\bibinfo {title} {Block-correlated
  coupled cluster theory: {The} general formulation and its application to the
  antiferromagnetic {Heisenberg} model},}\ }\href {\doibase 10.1063/1.1646355}
  {\bibfield  {journal} {\bibinfo  {journal} {J. Chem. Phys.}\ }\textbf
  {\bibinfo {volume} {120}},\ \bibinfo {pages} {5017--5026} (\bibinfo {year}
  {2004})}\BibitemShut {NoStop}%
\bibitem [{\citenamefont {Wang}\ \emph {et~al.}(2020)\citenamefont {Wang},
  \citenamefont {Duan}, \citenamefont {Xu}, \citenamefont {Zou},\ and\
  \citenamefont {Li}}]{wang_describing_2020}%
  \BibitemOpen
  \bibfield  {author} {\bibinfo {author} {\bibfnamefont {Q.}~\bibnamefont
  {Wang}}, \bibinfo {author} {\bibfnamefont {M.}~\bibnamefont {Duan}}, \bibinfo
  {author} {\bibfnamefont {E.}~\bibnamefont {Xu}}, \bibinfo {author}
  {\bibfnamefont {J.}~\bibnamefont {Zou}}, \ and\ \bibinfo {author}
  {\bibfnamefont {S.}~\bibnamefont {Li}},\ }\bibfield  {title} {\enquote
  {\bibinfo {title} {Describing {Strong} {Correlation} with
  {Block}-{Correlated} {Coupled} {Cluster} {Theory}},}\ }\href {\doibase
  10.1021/acs.jpclett.0c02117} {\bibfield  {journal} {\bibinfo  {journal} {J.
  Phys. Chem. Lett.}\ }\textbf {\bibinfo {volume} {11}},\ \bibinfo {pages}
  {7536--7543} (\bibinfo {year} {2020})}\BibitemShut {NoStop}%
\bibitem [{\citenamefont {Abraham}\ and\ \citenamefont
  {Mayhall}(2020)}]{abraham_selected_2020}%
  \BibitemOpen
  \bibfield  {author} {\bibinfo {author} {\bibfnamefont {V.}~\bibnamefont
  {Abraham}}\ and\ \bibinfo {author} {\bibfnamefont {N.~J.}\ \bibnamefont
  {Mayhall}},\ }\bibfield  {title} {\enquote {\bibinfo {title} {Selected
  {Configuration} {Interaction} in a {Basis} of {Cluster} {State} {Tensor}
  {Products}},}\ }\href {\doibase 10.1021/acs.jctc.0c00141} {\bibfield
  {journal} {\bibinfo  {journal} {J. Chem. Theory Comput.}\ }\textbf {\bibinfo
  {volume} {16}},\ \bibinfo {pages} {6098--6113} (\bibinfo {year}
  {2020})}\BibitemShut {NoStop}%
\bibitem [{\citenamefont {Knizia}\ and\ \citenamefont
  {Chan}(2012)}]{knizia_density_2012}%
  \BibitemOpen
  \bibfield  {author} {\bibinfo {author} {\bibfnamefont {G.}~\bibnamefont
  {Knizia}}\ and\ \bibinfo {author} {\bibfnamefont {G.~K.-L.}\ \bibnamefont
  {Chan}},\ }\bibfield  {title} {\enquote {\bibinfo {title} {Density {Matrix}
  {Embedding}: {A} {Simple} {Alternative} to {Dynamical} {Mean}-{Field}
  {Theory}},}\ }\href {\doibase 10.1103/PhysRevLett.109.186404} {\bibfield
  {journal} {\bibinfo  {journal} {Phys. Rev. Lett.}\ }\textbf {\bibinfo
  {volume} {109}},\ \bibinfo {pages} {186404} (\bibinfo {year}
  {2012})}\BibitemShut {NoStop}%
\bibitem [{\citenamefont {Wouters}\ \emph {et~al.}(2016)\citenamefont
  {Wouters}, \citenamefont {Jiménez-Hoyos}, \citenamefont {Sun},\ and\
  \citenamefont {Chan}}]{wouters_practical_2016}%
  \BibitemOpen
  \bibfield  {author} {\bibinfo {author} {\bibfnamefont {S.}~\bibnamefont
  {Wouters}}, \bibinfo {author} {\bibfnamefont {C.~A.}\ \bibnamefont
  {Jiménez-Hoyos}}, \bibinfo {author} {\bibfnamefont {Q.}~\bibnamefont {Sun}},
  \ and\ \bibinfo {author} {\bibfnamefont {G.~K.-L.}\ \bibnamefont {Chan}},\
  }\bibfield  {title} {\enquote {\bibinfo {title} {A {Practical} {Guide} to
  {Density} {Matrix} {Embedding} {Theory} in {Quantum} {Chemistry}},}\ }\href
  {\doibase 10.1021/acs.jctc.6b00316} {\bibfield  {journal} {\bibinfo
  {journal} {J. Chem. Theory Comput.}\ }\textbf {\bibinfo {volume} {12}},\
  \bibinfo {pages} {2706--2719} (\bibinfo {year} {2016})}\BibitemShut {NoStop}%
\bibitem [{\citenamefont {Parker}\ \emph {et~al.}(2013)\citenamefont {Parker},
  \citenamefont {Seideman}, \citenamefont {Ratner},\ and\ \citenamefont
  {Shiozaki}}]{parker_communication_2013}%
  \BibitemOpen
  \bibfield  {author} {\bibinfo {author} {\bibfnamefont {S.~M.}\ \bibnamefont
  {Parker}}, \bibinfo {author} {\bibfnamefont {T.}~\bibnamefont {Seideman}},
  \bibinfo {author} {\bibfnamefont {M.~A.}\ \bibnamefont {Ratner}}, \ and\
  \bibinfo {author} {\bibfnamefont {T.}~\bibnamefont {Shiozaki}},\ }\bibfield
  {title} {\enquote {\bibinfo {title} {Communication: {Active}-space
  decomposition for molecular dimers},}\ }\href {\doibase 10.1063/1.4813827}
  {\bibfield  {journal} {\bibinfo  {journal} {J. Chem. Phys.}\ }\textbf
  {\bibinfo {volume} {139}},\ \bibinfo {pages} {021108} (\bibinfo {year}
  {2013})}\BibitemShut {NoStop}%
\bibitem [{\citenamefont {Hermes}\ and\ \citenamefont
  {Gagliardi}(2019)}]{hermes_multiconfigurational_2019}%
  \BibitemOpen
  \bibfield  {author} {\bibinfo {author} {\bibfnamefont {M.~R.}\ \bibnamefont
  {Hermes}}\ and\ \bibinfo {author} {\bibfnamefont {L.}~\bibnamefont
  {Gagliardi}},\ }\bibfield  {title} {\enquote {\bibinfo {title}
  {Multiconfigurational {Self}-{Consistent} {Field} {Theory} with {Density}
  {Matrix} {Embedding}: {The} {Localized} {Active} {Space} {Self}-{Consistent}
  {Field} {Method}},}\ }\href {\doibase 10.1021/acs.jctc.8b01009} {\bibfield
  {journal} {\bibinfo  {journal} {J. Chem. Theory Comput.}\ }\textbf {\bibinfo
  {volume} {15}},\ \bibinfo {pages} {972--986} (\bibinfo {year}
  {2019})}\BibitemShut {NoStop}%
\bibitem [{\citenamefont {Isaev}, \citenamefont {Ortiz},\ and\ \citenamefont
  {Dukelsky}(2009)}]{isaev_hierarchical_2009}%
  \BibitemOpen
  \bibfield  {author} {\bibinfo {author} {\bibfnamefont {L.}~\bibnamefont
  {Isaev}}, \bibinfo {author} {\bibfnamefont {G.}~\bibnamefont {Ortiz}}, \ and\
  \bibinfo {author} {\bibfnamefont {J.}~\bibnamefont {Dukelsky}},\ }\bibfield
  {title} {\enquote {\bibinfo {title} {Hierarchical mean-field approach to the
  {J}$_{\textrm{1}}$ −{J}$_{\textrm{2}}$ heisenberg model on a square
  lattice},}\ }\href {\doibase 10.1103/PhysRevB.79.024409} {\bibfield
  {journal} {\bibinfo  {journal} {Phys. Rev. B}\ }\textbf {\bibinfo {volume}
  {79}},\ \bibinfo {pages} {024409} (\bibinfo {year} {2009})}\BibitemShut
  {NoStop}%
\bibitem [{\citenamefont {Ghassemi~Tabrizi}\ and\ \citenamefont
  {Jiménez-Hoyos}()}]{ghassemi_tabrizi_ground_2023}%
  \BibitemOpen
  \bibfield  {author} {\bibinfo {author} {\bibfnamefont {S.}~\bibnamefont
  {Ghassemi~Tabrizi}}\ and\ \bibinfo {author} {\bibfnamefont {C.~A.}\
  \bibnamefont {Jiménez-Hoyos}},\ }\bibfield  {title} {\enquote {\bibinfo
  {title} {Ground states of heisenberg spin clusters from a cluster-based
  projected hartree–fock approach},}\ }\href {\doibase
  10.3390/condmat8010018} {\ \textbf {\bibinfo {volume} {8}},\ \bibinfo {pages}
  {18}}\BibitemShut {NoStop}%
\bibitem [{\citenamefont {Fang}\ and\ \citenamefont
  {Li}(2007)}]{fang_block_2007}%
  \BibitemOpen
  \bibfield  {author} {\bibinfo {author} {\bibfnamefont {T.}~\bibnamefont
  {Fang}}\ and\ \bibinfo {author} {\bibfnamefont {S.}~\bibnamefont {Li}},\
  }\bibfield  {title} {\enquote {\bibinfo {title} {Block correlated coupled
  cluster theory with a complete active-space self-consistent-field reference
  function: {The} formulation and test applications for single bond
  breaking},}\ }\href {\doibase 10.1063/1.2800027} {\bibfield  {journal}
  {\bibinfo  {journal} {J. Chem. Phys.}\ }\textbf {\bibinfo {volume} {127}},\
  \bibinfo {pages} {204108} (\bibinfo {year} {2007})}\BibitemShut {NoStop}%
\bibitem [{\citenamefont {Nakatani}(2018)}]{nakatani_matrix_2018}%
  \BibitemOpen
  \bibfield  {author} {\bibinfo {author} {\bibfnamefont {N.}~\bibnamefont
  {Nakatani}},\ }\bibfield  {title} {\enquote {\bibinfo {title} {Matrix
  {Product} {States} and {Density} {Matrix} {Renormalization} {Group}
  {Algorithm}},}\ }in\ \href {\doibase 10.1016/B978-0-12-409547-2.11473-8}
  {\emph {\bibinfo {booktitle} {Reference {Module} in {Chemistry}, {Molecular}
  {Sciences} and {Chemical} {Engineering}}}}\ (\bibinfo  {publisher}
  {Elsevier},\ \bibinfo {year} {2018})\BibitemShut {NoStop}%
\bibitem [{\citenamefont {Sharma}\ \emph {et~al.}(2014)\citenamefont {Sharma},
  \citenamefont {Sivalingam}, \citenamefont {Neese},\ and\ \citenamefont
  {Chan}}]{chan}%
  \BibitemOpen
  \bibfield  {author} {\bibinfo {author} {\bibfnamefont {S.}~\bibnamefont
  {Sharma}}, \bibinfo {author} {\bibfnamefont {K.}~\bibnamefont {Sivalingam}},
  \bibinfo {author} {\bibfnamefont {F.}~\bibnamefont {Neese}}, \ and\ \bibinfo
  {author} {\bibfnamefont {G.~K.-L.}\ \bibnamefont {Chan}},\ }\bibfield
  {title} {\enquote {\bibinfo {title} {Low-energy spectrum of iron–sulfur
  clusters directly from many-particle quantum mechanics},}\ }\href {\doibase
  10.1038/nchem.2041} {\bibfield  {journal} {\bibinfo  {journal} {Nature Chem}\
  }\textbf {\bibinfo {volume} {6}},\ \bibinfo {pages} {927--933} (\bibinfo
  {year} {2014})}\BibitemShut {NoStop}%
\bibitem [{\citenamefont {Rams}\ \emph {et~al.}(2020)\citenamefont {Rams},
  \citenamefont {Jochim}, \citenamefont {Böhme}, \citenamefont {Lohmiller},
  \citenamefont {Ceglarska}, \citenamefont {Rams}, \citenamefont {Schnegg},
  \citenamefont {Plass},\ and\ \citenamefont
  {Näther}}]{rams_singlechain_2020}%
  \BibitemOpen
  \bibfield  {author} {\bibinfo {author} {\bibfnamefont {M.}~\bibnamefont
  {Rams}}, \bibinfo {author} {\bibfnamefont {A.}~\bibnamefont {Jochim}},
  \bibinfo {author} {\bibfnamefont {M.}~\bibnamefont {Böhme}}, \bibinfo
  {author} {\bibfnamefont {T.}~\bibnamefont {Lohmiller}}, \bibinfo {author}
  {\bibfnamefont {M.}~\bibnamefont {Ceglarska}}, \bibinfo {author}
  {\bibfnamefont {M.~M.}\ \bibnamefont {Rams}}, \bibinfo {author}
  {\bibfnamefont {A.}~\bibnamefont {Schnegg}}, \bibinfo {author} {\bibfnamefont
  {W.}~\bibnamefont {Plass}}, \ and\ \bibinfo {author} {\bibfnamefont
  {C.}~\bibnamefont {Näther}},\ }\bibfield  {title} {\enquote {\bibinfo
  {title} {Single‐{Chain} {Magnet} {Based} on {Cobalt}({II}) {Thiocyanate} as
  {XXZ} {Spin} {Chain}},}\ }\href {\doibase 10.1002/chem.201903924} {\bibfield
  {journal} {\bibinfo  {journal} {Chem. Eur. J.}\ }\textbf {\bibinfo {volume}
  {26}},\ \bibinfo {pages} {2837--2851} (\bibinfo {year} {2020})}\BibitemShut
  {NoStop}%
\bibitem [{\citenamefont {Wu}, \citenamefont {Schmalz},\ and\ \citenamefont
  {Klein}(2002)}]{apps1}%
  \BibitemOpen
  \bibfield  {author} {\bibinfo {author} {\bibfnamefont {J.}~\bibnamefont
  {Wu}}, \bibinfo {author} {\bibfnamefont {T.~G.}\ \bibnamefont {Schmalz}}, \
  and\ \bibinfo {author} {\bibfnamefont {D.~J.}\ \bibnamefont {Klein}},\
  }\bibfield  {title} {\enquote {\bibinfo {title} {An extended {Heisenberg}
  model for conjugated hydrocarbons},}\ }\href {\doibase 10.1063/1.1520133}
  {\bibfield  {journal} {\bibinfo  {journal} {J. Chem. Phys.}\ }\textbf
  {\bibinfo {volume} {117}},\ \bibinfo {pages} {9977--9982} (\bibinfo {year}
  {2002})}\BibitemShut {NoStop}%
\bibitem [{\citenamefont {Dye}(1997)}]{apps2}%
  \BibitemOpen
  \bibfield  {author} {\bibinfo {author} {\bibfnamefont {J.~L.}\ \bibnamefont
  {Dye}},\ }\bibfield  {title} {\enquote {\bibinfo {title} {Electrides: {From}
  {1D} {Heisenberg} {Chains} to {2D} {Pseudo}-{Metals}},}\ }\href {\doibase
  10.1021/ic970551z} {\bibfield  {journal} {\bibinfo  {journal} {Inorg. Chem.}\
  }\textbf {\bibinfo {volume} {36}},\ \bibinfo {pages} {3816--3826} (\bibinfo
  {year} {1997})}\BibitemShut {NoStop}%
\bibitem [{\citenamefont {Van~Oosten}, \citenamefont {Broer},\ and\
  \citenamefont {Nieuwpoort}(1995)}]{apps3}%
  \BibitemOpen
  \bibfield  {author} {\bibinfo {author} {\bibfnamefont {A.~B.}\ \bibnamefont
  {Van~Oosten}}, \bibinfo {author} {\bibfnamefont {R.}~\bibnamefont {Broer}}, \
  and\ \bibinfo {author} {\bibfnamefont {W.~C.}\ \bibnamefont {Nieuwpoort}},\
  }\bibfield  {title} {\enquote {\bibinfo {title} {Heisenberg exchange in
  {La$_{\textrm{2}}$CuO$_{\textrm{4}}$}},}\ }\href {\doibase
  10.1002/qua.560560826} {\bibfield  {journal} {\bibinfo  {journal} {Int. J.
  Quantum Chem.}\ }\textbf {\bibinfo {volume} {56}},\ \bibinfo {pages}
  {241--243} (\bibinfo {year} {1995})}\BibitemShut {NoStop}%
\bibitem [{\citenamefont {Bethe}(1931)}]{bethe_zur_1931}%
  \BibitemOpen
  \bibfield  {author} {\bibinfo {author} {\bibfnamefont {H.}~\bibnamefont
  {Bethe}},\ }\bibfield  {title} {\enquote {\bibinfo {title} {Zur {Theorie} der
  {Metalle}: {I}. {Eigenwerte} und {Eigenfunktionen} der linearen
  {Atomkette}},}\ }\href {\doibase 10.1007/BF01341708} {\bibfield  {journal}
  {\bibinfo  {journal} {Z. Phys.}\ }\textbf {\bibinfo {volume} {71}},\ \bibinfo
  {pages} {205--226} (\bibinfo {year} {1931})}\BibitemShut {NoStop}%
\bibitem [{\citenamefont {Yang}\ and\ \citenamefont
  {Yang}(1966)}]{yang_ground-state_1966}%
  \BibitemOpen
  \bibfield  {author} {\bibinfo {author} {\bibfnamefont {C.~N.}\ \bibnamefont
  {Yang}}\ and\ \bibinfo {author} {\bibfnamefont {C.~P.}\ \bibnamefont
  {Yang}},\ }\bibfield  {title} {\enquote {\bibinfo {title} {Ground-{State}
  {Energy} of a {Heisenberg}-{Ising} {Lattice}},}\ }\href {\doibase
  10.1103/PhysRev.147.303} {\bibfield  {journal} {\bibinfo  {journal} {Phys.
  Rev.}\ }\textbf {\bibinfo {volume} {147}},\ \bibinfo {pages} {303--306}
  (\bibinfo {year} {1966})}\BibitemShut {NoStop}%
\bibitem [{\citenamefont {Dagotto}\ and\ \citenamefont
  {Moreo}(1989)}]{dagotto_phase_1989}%
  \BibitemOpen
  \bibfield  {author} {\bibinfo {author} {\bibfnamefont {E.}~\bibnamefont
  {Dagotto}}\ and\ \bibinfo {author} {\bibfnamefont {A.}~\bibnamefont
  {Moreo}},\ }\bibfield  {title} {\enquote {\bibinfo {title} {Phase diagram of
  the frustrated spin-1/2 {Heisenberg} antiferromagnet in 2 dimensions},}\
  }\href {\doibase 10.1103/PhysRevLett.63.2148} {\bibfield  {journal} {\bibinfo
   {journal} {Phys. Rev. Lett.}\ }\textbf {\bibinfo {volume} {63}},\ \bibinfo
  {pages} {2148--2151} (\bibinfo {year} {1989})}\BibitemShut {NoStop}%
\bibitem [{\citenamefont {Schulz}\ and\ \citenamefont
  {Ziman}(1992)}]{schulz_finite-size_1992}%
  \BibitemOpen
  \bibfield  {author} {\bibinfo {author} {\bibfnamefont {H.~J.}\ \bibnamefont
  {Schulz}}\ and\ \bibinfo {author} {\bibfnamefont {T.~A.~L.}\ \bibnamefont
  {Ziman}},\ }\bibfield  {title} {\enquote {\bibinfo {title} {Finite-{Size}
  {Scaling} for the {Two}-{Dimensional} {Frustrated} {Quantum} {Heisenberg}
  {Antiferromagnet}},}\ }\href {\doibase 10.1209/0295-5075/18/4/013} {\bibfield
   {journal} {\bibinfo  {journal} {Europhys. Lett.}\ }\textbf {\bibinfo
  {volume} {18}},\ \bibinfo {pages} {355--360} (\bibinfo {year}
  {1992})}\BibitemShut {NoStop}%
\bibitem [{\citenamefont {Richter}\ and\ \citenamefont
  {Schulenburg}(2010)}]{richter_spin-12_2010}%
  \BibitemOpen
  \bibfield  {author} {\bibinfo {author} {\bibfnamefont {J.}~\bibnamefont
  {Richter}}\ and\ \bibinfo {author} {\bibfnamefont {J.}~\bibnamefont
  {Schulenburg}},\ }\bibfield  {title} {\enquote {\bibinfo {title} {The
  spin-1/2 {J}$_{\textrm{1}}$−{J}$_{\textrm{2}}$ {Heisenberg} antiferromagnet
  on the square lattice: {Exact} diagonalization for {N}=40 spins},}\ }\href
  {\doibase 10.1140/epjb/e2009-00400-4} {\bibfield  {journal} {\bibinfo
  {journal} {Eur. Phys. J. B}\ }\textbf {\bibinfo {volume} {73}},\ \bibinfo
  {pages} {117--124} (\bibinfo {year} {2010})}\BibitemShut {NoStop}%
\bibitem [{\citenamefont {Capriotti}\ and\ \citenamefont
  {Sorella}(2000)}]{capriotti_spontaneous_2000}%
  \BibitemOpen
  \bibfield  {author} {\bibinfo {author} {\bibfnamefont {L.}~\bibnamefont
  {Capriotti}}\ and\ \bibinfo {author} {\bibfnamefont {S.}~\bibnamefont
  {Sorella}},\ }\bibfield  {title} {\enquote {\bibinfo {title} {Spontaneous
  {Plaquette} {Dimerization} in the {J}$_{\textrm{1}}$−{J}$_{\textrm{2}}$
  {Heisenberg} {Model}},}\ }\href {\doibase 10.1103/PhysRevLett.84.3173}
  {\bibfield  {journal} {\bibinfo  {journal} {Phys. Rev. Lett.}\ }\textbf
  {\bibinfo {volume} {84}},\ \bibinfo {pages} {3173--3176} (\bibinfo {year}
  {2000})}\BibitemShut {NoStop}%
\bibitem [{\citenamefont {Mambrini}\ \emph {et~al.}(2006)\citenamefont
  {Mambrini}, \citenamefont {Läuchli}, \citenamefont {Poilblanc},\ and\
  \citenamefont {Mila}}]{mambrini_plaquette_2006}%
  \BibitemOpen
  \bibfield  {author} {\bibinfo {author} {\bibfnamefont {M.}~\bibnamefont
  {Mambrini}}, \bibinfo {author} {\bibfnamefont {A.}~\bibnamefont {Läuchli}},
  \bibinfo {author} {\bibfnamefont {D.}~\bibnamefont {Poilblanc}}, \ and\
  \bibinfo {author} {\bibfnamefont {F.}~\bibnamefont {Mila}},\ }\bibfield
  {title} {\enquote {\bibinfo {title} {Plaquette valence-bond crystal in the
  frustrated {Heisenberg} quantum antiferromagnet on the square lattice},}\
  }\href {\doibase 10.1103/PhysRevB.74.144422} {\bibfield  {journal} {\bibinfo
  {journal} {Phys. Rev. B}\ }\textbf {\bibinfo {volume} {74}},\ \bibinfo
  {pages} {144422} (\bibinfo {year} {2006})}\BibitemShut {NoStop}%
\bibitem [{\citenamefont {Schulz}, \citenamefont {Ziman},\ and\ \citenamefont
  {Poilblanc}(1996)}]{schulz_magnetic_1996}%
  \BibitemOpen
  \bibfield  {author} {\bibinfo {author} {\bibfnamefont {H.~J.}\ \bibnamefont
  {Schulz}}, \bibinfo {author} {\bibfnamefont {T.~A.}\ \bibnamefont {Ziman}}, \
  and\ \bibinfo {author} {\bibfnamefont {D.}~\bibnamefont {Poilblanc}},\
  }\bibfield  {title} {\enquote {\bibinfo {title} {Magnetic {Order} and
  {Disorder} in the {Frustrated} {Quantum} {Heisenberg} {Antiferromagnet} in
  {Two} {Dimensions}},}\ }\href {\doibase 10.1051/jp1:1996236} {\bibfield
  {journal} {\bibinfo  {journal} {J. Phys., I}\ }\textbf {\bibinfo {volume}
  {6}},\ \bibinfo {pages} {675--703} (\bibinfo {year} {1996})}\BibitemShut
  {NoStop}%
\bibitem [{\citenamefont {Schmalfuß}\ \emph {et~al.}(2006)\citenamefont
  {Schmalfuß}, \citenamefont {Darradi}, \citenamefont {Richter}, \citenamefont
  {Schulenburg},\ and\ \citenamefont {Ihle}}]{schmalfus_quantum_2006}%
  \BibitemOpen
  \bibfield  {author} {\bibinfo {author} {\bibfnamefont {D.}~\bibnamefont
  {Schmalfuß}}, \bibinfo {author} {\bibfnamefont {R.}~\bibnamefont {Darradi}},
  \bibinfo {author} {\bibfnamefont {J.}~\bibnamefont {Richter}}, \bibinfo
  {author} {\bibfnamefont {J.}~\bibnamefont {Schulenburg}}, \ and\ \bibinfo
  {author} {\bibfnamefont {D.}~\bibnamefont {Ihle}},\ }\bibfield  {title}
  {\enquote {\bibinfo {title} {Quantum {J} 1 − {J} 2 {Antiferromagnet} on a
  {Stacked} {Square} {Lattice}: {Influence} of the {Interlayer} {Coupling} on
  the {Ground}-{State} {Magnetic} {Ordering}},}\ }\href {\doibase
  10.1103/PhysRevLett.97.157201} {\bibfield  {journal} {\bibinfo  {journal}
  {Phys. Rev. Lett.}\ }\textbf {\bibinfo {volume} {97}},\ \bibinfo {pages}
  {157201} (\bibinfo {year} {2006})}\BibitemShut {NoStop}%
\bibitem [{\citenamefont {Darradi}\ \emph {et~al.}(2008)\citenamefont
  {Darradi}, \citenamefont {Derzhko}, \citenamefont {Zinke}, \citenamefont
  {Schulenburg}, \citenamefont {Krüger},\ and\ \citenamefont
  {Richter}}]{darradi_ground_2008}%
  \BibitemOpen
  \bibfield  {author} {\bibinfo {author} {\bibfnamefont {R.}~\bibnamefont
  {Darradi}}, \bibinfo {author} {\bibfnamefont {O.}~\bibnamefont {Derzhko}},
  \bibinfo {author} {\bibfnamefont {R.}~\bibnamefont {Zinke}}, \bibinfo
  {author} {\bibfnamefont {J.}~\bibnamefont {Schulenburg}}, \bibinfo {author}
  {\bibfnamefont {S.~E.}\ \bibnamefont {Krüger}}, \ and\ \bibinfo {author}
  {\bibfnamefont {J.}~\bibnamefont {Richter}},\ }\bibfield  {title} {\enquote
  {\bibinfo {title} {Ground state phases of the spin-1/2
  {J}$_{\textrm{1}}$−{J}$_{\textrm{2}}$ {Heisenberg} antiferromagnet on the
  square lattice: {A} high-order coupled cluster treatment},}\ }\href {\doibase
  10.1103/PhysRevB.78.214415} {\bibfield  {journal} {\bibinfo  {journal} {Phys.
  Rev. B}\ }\textbf {\bibinfo {volume} {78}},\ \bibinfo {pages} {214415}
  (\bibinfo {year} {2008})}\BibitemShut {NoStop}%
\bibitem [{\citenamefont {Bishop}, \citenamefont {Farnell},\ and\ \citenamefont
  {Parkinson}(1998)}]{bishop_phase_1998}%
  \BibitemOpen
  \bibfield  {author} {\bibinfo {author} {\bibfnamefont {R.~F.}\ \bibnamefont
  {Bishop}}, \bibinfo {author} {\bibfnamefont {D.~J.~J.}\ \bibnamefont
  {Farnell}}, \ and\ \bibinfo {author} {\bibfnamefont {J.~B.}\ \bibnamefont
  {Parkinson}},\ }\bibfield  {title} {\enquote {\bibinfo {title} {Phase
  transitions in the spin-half {J} 1 − {J} 2 model},}\ }\href {\doibase
  10.1103/PhysRevB.58.6394} {\bibfield  {journal} {\bibinfo  {journal} {Phys.
  Rev. B}\ }\textbf {\bibinfo {volume} {58}},\ \bibinfo {pages} {6394--6402}
  (\bibinfo {year} {1998})}\BibitemShut {NoStop}%
\bibitem [{\citenamefont {Richter}, \citenamefont {Zinke},\ and\ \citenamefont
  {Farnell}(2015)}]{richter_spin-12_2015}%
  \BibitemOpen
  \bibfield  {author} {\bibinfo {author} {\bibfnamefont {J.}~\bibnamefont
  {Richter}}, \bibinfo {author} {\bibfnamefont {R.}~\bibnamefont {Zinke}}, \
  and\ \bibinfo {author} {\bibfnamefont {D.~J.~J.}\ \bibnamefont {Farnell}},\
  }\bibfield  {title} {\enquote {\bibinfo {title} {The spin-1/2 square-lattice
  {J$_{\textrm{1}}$}-{J$_{\textrm{2}}$} model: the spin-gap issue},}\ }\href
  {\doibase 10.1140/epjb/e2014-50589-x} {\bibfield  {journal} {\bibinfo
  {journal} {Eur. Phys. J. B}\ }\textbf {\bibinfo {volume} {88}},\ \bibinfo
  {pages} {2} (\bibinfo {year} {2015})}\BibitemShut {NoStop}%
\bibitem [{\citenamefont {Farnell}\ \emph {et~al.}(2009)\citenamefont
  {Farnell}, \citenamefont {Richter}, \citenamefont {Zinke},\ and\
  \citenamefont {Bishop}}]{bishop_main}%
  \BibitemOpen
  \bibfield  {author} {\bibinfo {author} {\bibfnamefont {D.~J.~J.}\
  \bibnamefont {Farnell}}, \bibinfo {author} {\bibfnamefont {J.}~\bibnamefont
  {Richter}}, \bibinfo {author} {\bibfnamefont {R.}~\bibnamefont {Zinke}}, \
  and\ \bibinfo {author} {\bibfnamefont {R.~F.}\ \bibnamefont {Bishop}},\
  }\bibfield  {title} {\enquote {\bibinfo {title} {High-order coupled cluster
  method (ccm) calculations for quantum magnets with valence-bond ground
  states},}\ }\href {\doibase 10.1007/s10955-009-9703-7} {\bibfield  {journal}
  {\bibinfo  {journal} {J. Stat. Phys.}\ }\textbf {\bibinfo {volume} {135}}
  (\bibinfo {year} {2009}),\ 10.1007/s10955-009-9703-7}\BibitemShut {NoStop}%
\bibitem [{\citenamefont {Jiang}, \citenamefont {Yao},\ and\ \citenamefont
  {Balents}(2012)}]{jiang_spin_2012}%
  \BibitemOpen
  \bibfield  {author} {\bibinfo {author} {\bibfnamefont {H.-C.}\ \bibnamefont
  {Jiang}}, \bibinfo {author} {\bibfnamefont {H.}~\bibnamefont {Yao}}, \ and\
  \bibinfo {author} {\bibfnamefont {L.}~\bibnamefont {Balents}},\ }\bibfield
  {title} {\enquote {\bibinfo {title} {Spin liquid ground state of the spin-1/2
  square {J}$_{\textrm{1}}$−{J}$_{\textrm{2}}$ {Heisenberg} model},}\ }\href
  {\doibase 10.1103/PhysRevB.86.024424} {\bibfield  {journal} {\bibinfo
  {journal} {Phys. Rev. B}\ }\textbf {\bibinfo {volume} {86}},\ \bibinfo
  {pages} {024424} (\bibinfo {year} {2012})}\BibitemShut {NoStop}%
\bibitem [{\citenamefont {Gong}\ \emph {et~al.}(2014)\citenamefont {Gong},
  \citenamefont {Zhu}, \citenamefont {Sheng}, \citenamefont {Motrunich},\ and\
  \citenamefont {Fisher}}]{gong_plaquette_2014}%
  \BibitemOpen
  \bibfield  {author} {\bibinfo {author} {\bibfnamefont {S.-S.}\ \bibnamefont
  {Gong}}, \bibinfo {author} {\bibfnamefont {W.}~\bibnamefont {Zhu}}, \bibinfo
  {author} {\bibfnamefont {D.}~\bibnamefont {Sheng}}, \bibinfo {author}
  {\bibfnamefont {O.~I.}\ \bibnamefont {Motrunich}}, \ and\ \bibinfo {author}
  {\bibfnamefont {M.~P.}\ \bibnamefont {Fisher}},\ }\bibfield  {title}
  {\enquote {\bibinfo {title} {Plaquette {Ordered} {Phase} and {Quantum}
  {Phase} {Diagram} in the {Spin}-1/2 {J}$_{\textrm{1}}$−{J}$_{\textrm{2}}$
  {Square} {Heisenberg} {Model}},}\ }\href {\doibase
  10.1103/PhysRevLett.113.027201} {\bibfield  {journal} {\bibinfo  {journal}
  {Phys. Rev. Lett.}\ }\textbf {\bibinfo {volume} {113}},\ \bibinfo {pages}
  {027201} (\bibinfo {year} {2014})}\BibitemShut {NoStop}%
\bibitem [{\citenamefont {Murg}, \citenamefont {Verstraete},\ and\
  \citenamefont {Cirac}(2009)}]{murg_exploring_2009}%
  \BibitemOpen
  \bibfield  {author} {\bibinfo {author} {\bibfnamefont {V.}~\bibnamefont
  {Murg}}, \bibinfo {author} {\bibfnamefont {F.}~\bibnamefont {Verstraete}}, \
  and\ \bibinfo {author} {\bibfnamefont {J.~I.}\ \bibnamefont {Cirac}},\
  }\bibfield  {title} {\enquote {\bibinfo {title} {Exploring frustrated spin
  systems using projected entangled pair states},}\ }\href {\doibase
  10.1103/PhysRevB.79.195119} {\bibfield  {journal} {\bibinfo  {journal} {Phys.
  Rev. B}\ }\textbf {\bibinfo {volume} {79}},\ \bibinfo {pages} {195119}
  (\bibinfo {year} {2009})}\BibitemShut {NoStop}%
\bibitem [{\citenamefont {Yu}\ and\ \citenamefont {Kao}(2012)}]{yu_spin-_2012}%
  \BibitemOpen
  \bibfield  {author} {\bibinfo {author} {\bibfnamefont {J.-F.}\ \bibnamefont
  {Yu}}\ and\ \bibinfo {author} {\bibfnamefont {Y.-J.}\ \bibnamefont {Kao}},\
  }\bibfield  {title} {\enquote {\bibinfo {title} {Spin-1/2
  {J}$_{\textrm{1}}$−{J}$_{\textrm{2}}$ {Heisenberg} antiferromagnet on a
  square lattice: {A} plaquette renormalized tensor network study},}\ }\href
  {\doibase 10.1103/PhysRevB.85.094407} {\bibfield  {journal} {\bibinfo
  {journal} {Phys. Rev. B}\ }\textbf {\bibinfo {volume} {85}},\ \bibinfo
  {pages} {094407} (\bibinfo {year} {2012})}\BibitemShut {NoStop}%
\bibitem [{\citenamefont {Wang}\ \emph {et~al.}(2013)\citenamefont {Wang},
  \citenamefont {Poilblanc}, \citenamefont {Gu}, \citenamefont {Wen},\ and\
  \citenamefont {Verstraete}}]{wang_constructing_2013}%
  \BibitemOpen
  \bibfield  {author} {\bibinfo {author} {\bibfnamefont {L.}~\bibnamefont
  {Wang}}, \bibinfo {author} {\bibfnamefont {D.}~\bibnamefont {Poilblanc}},
  \bibinfo {author} {\bibfnamefont {Z.-C.}\ \bibnamefont {Gu}}, \bibinfo
  {author} {\bibfnamefont {X.-G.}\ \bibnamefont {Wen}}, \ and\ \bibinfo
  {author} {\bibfnamefont {F.}~\bibnamefont {Verstraete}},\ }\bibfield  {title}
  {\enquote {\bibinfo {title} {Constructing a {Gapless} {Spin}-{Liquid} {State}
  for the {Spin}-1/2 {J}$_{\textrm{1}}$−{J}$_{\textrm{2}}$ {Heisenberg}
  {Model} on a {Square} {Lattice}},}\ }\href {\doibase
  10.1103/PhysRevLett.111.037202} {\bibfield  {journal} {\bibinfo  {journal}
  {Phys. Rev. Lett.}\ }\textbf {\bibinfo {volume} {111}},\ \bibinfo {pages}
  {037202} (\bibinfo {year} {2013})}\BibitemShut {NoStop}%
\bibitem [{\citenamefont {Capriotti}\ \emph {et~al.}(2001)\citenamefont
  {Capriotti}, \citenamefont {Becca}, \citenamefont {Parola},\ and\
  \citenamefont {Sorella}}]{capriotti_resonating_2001}%
  \BibitemOpen
  \bibfield  {author} {\bibinfo {author} {\bibfnamefont {L.}~\bibnamefont
  {Capriotti}}, \bibinfo {author} {\bibfnamefont {F.}~\bibnamefont {Becca}},
  \bibinfo {author} {\bibfnamefont {A.}~\bibnamefont {Parola}}, \ and\ \bibinfo
  {author} {\bibfnamefont {S.}~\bibnamefont {Sorella}},\ }\bibfield  {title}
  {\enquote {\bibinfo {title} {Resonating {Valence} {Bond} {Wave} {Functions}
  for {Strongly} {Frustrated} {Spin} {Systems}},}\ }\href {\doibase
  10.1103/PhysRevLett.87.097201} {\bibfield  {journal} {\bibinfo  {journal}
  {Phys. Rev. Lett.}\ }\textbf {\bibinfo {volume} {87}},\ \bibinfo {pages}
  {097201} (\bibinfo {year} {2001})}\BibitemShut {NoStop}%
\bibitem [{\citenamefont {Sandvik}(1997)}]{sandvik_finite-size_1997}%
  \BibitemOpen
  \bibfield  {author} {\bibinfo {author} {\bibfnamefont {A.~W.}\ \bibnamefont
  {Sandvik}},\ }\bibfield  {title} {\enquote {\bibinfo {title} {Finite-size
  scaling of the ground-state parameters of the two-dimensional {Heisenberg}
  model},}\ }\href {\doibase 10.1103/PhysRevB.56.11678} {\bibfield  {journal}
  {\bibinfo  {journal} {Phys. Rev. B}\ }\textbf {\bibinfo {volume} {56}},\
  \bibinfo {pages} {11678--11690} (\bibinfo {year} {1997})}\BibitemShut
  {NoStop}%
\bibitem [{\citenamefont {Massaccesi}\ \emph {et~al.}(2021)\citenamefont
  {Massaccesi}, \citenamefont {Rubio-García}, \citenamefont {Capuzzi},
  \citenamefont {Ríos}, \citenamefont {Oña}, \citenamefont {Dukelsky},
  \citenamefont {Lain}, \citenamefont {Torre},\ and\ \citenamefont
  {Alcoba}}]{massaccesi_variational_2021}%
  \BibitemOpen
  \bibfield  {author} {\bibinfo {author} {\bibfnamefont {G.~E.}\ \bibnamefont
  {Massaccesi}}, \bibinfo {author} {\bibfnamefont {A.}~\bibnamefont
  {Rubio-García}}, \bibinfo {author} {\bibfnamefont {P.}~\bibnamefont
  {Capuzzi}}, \bibinfo {author} {\bibfnamefont {E.}~\bibnamefont {Ríos}},
  \bibinfo {author} {\bibfnamefont {O.~B.}\ \bibnamefont {Oña}}, \bibinfo
  {author} {\bibfnamefont {J.}~\bibnamefont {Dukelsky}}, \bibinfo {author}
  {\bibfnamefont {L.}~\bibnamefont {Lain}}, \bibinfo {author} {\bibfnamefont
  {A.}~\bibnamefont {Torre}}, \ and\ \bibinfo {author} {\bibfnamefont {D.~R.}\
  \bibnamefont {Alcoba}},\ }\bibfield  {title} {\enquote {\bibinfo {title}
  {Variational determination of the two-particle reduced density matrix within
  the doubly occupied configuration interaction space: exploiting translational
  and reflection invariance},}\ }\href {\doibase 10.1088/1742-5468/abd940}
  {\bibfield  {journal} {\bibinfo  {journal} {J. Stat. Mech.}\ }\textbf
  {\bibinfo {volume} {2021}},\ \bibinfo {pages} {013110} (\bibinfo {year}
  {2021})}\BibitemShut {NoStop}%
\bibitem [{\citenamefont {de~Sousa}\ \emph {et~al.}(2003)\citenamefont
  {de~Sousa}, \citenamefont {Branco}, \citenamefont {Boechat},\ and\
  \citenamefont {Cordeiro}}]{de_sousa_quantum_2003}%
  \BibitemOpen
  \bibfield  {author} {\bibinfo {author} {\bibfnamefont {J.}~\bibnamefont
  {de~Sousa}}, \bibinfo {author} {\bibfnamefont {N.}~\bibnamefont {Branco}},
  \bibinfo {author} {\bibfnamefont {B.}~\bibnamefont {Boechat}}, \ and\
  \bibinfo {author} {\bibfnamefont {C.}~\bibnamefont {Cordeiro}},\ }\bibfield
  {title} {\enquote {\bibinfo {title} {Quantum spin- two-dimensional {XXZ}
  model: an alternative quantum renormalization-group approach},}\ }\href
  {\doibase 10.1016/S0378-4371(03)00544-2} {\bibfield  {journal} {\bibinfo
  {journal} {Physica A}\ }\textbf {\bibinfo {volume} {328}},\ \bibinfo {pages}
  {167--173} (\bibinfo {year} {2003})}\BibitemShut {NoStop}%
\bibitem [{\citenamefont {Cuccoli}, \citenamefont {Tognetti},\ and\
  \citenamefont {Vaia}(1995)}]{cuccoli_two-dimensional_1995}%
  \BibitemOpen
  \bibfield  {author} {\bibinfo {author} {\bibfnamefont {A.}~\bibnamefont
  {Cuccoli}}, \bibinfo {author} {\bibfnamefont {V.}~\bibnamefont {Tognetti}}, \
  and\ \bibinfo {author} {\bibfnamefont {R.}~\bibnamefont {Vaia}},\ }\bibfield
  {title} {\enquote {\bibinfo {title} {Two-dimensional \textit{{XXZ}} model on
  a square lattice: {A} {Monte} {Carlo} simulation},}\ }\href {\doibase
  10.1103/PhysRevB.52.10221} {\bibfield  {journal} {\bibinfo  {journal} {Phys.
  Rev. B}\ }\textbf {\bibinfo {volume} {52}},\ \bibinfo {pages} {10221--10231}
  (\bibinfo {year} {1995})}\BibitemShut {NoStop}%
\bibitem [{\citenamefont {Jung}\ and\ \citenamefont
  {Noh}(2020)}]{jung_guide_2020}%
  \BibitemOpen
  \bibfield  {author} {\bibinfo {author} {\bibfnamefont {J.-H.}\ \bibnamefont
  {Jung}}\ and\ \bibinfo {author} {\bibfnamefont {J.~D.}\ \bibnamefont {Noh}},\
  }\bibfield  {title} {\enquote {\bibinfo {title} {Guide to {Exact}
  {Diagonalization} {Study} of {Quantum} {Thermalization}},}\ }\href {\doibase
  10.3938/jkps.76.670} {\bibfield  {journal} {\bibinfo  {journal} {J. Korean
  Phys. Soc.}\ }\textbf {\bibinfo {volume} {76}},\ \bibinfo {pages} {670--683}
  (\bibinfo {year} {2020})}\BibitemShut {NoStop}%
\bibitem [{\citenamefont {Macrì}\ \emph {et~al.}(2021)\citenamefont {Macrì},
  \citenamefont {Lepori}, \citenamefont {Pagano}, \citenamefont {Lewenstein},\
  and\ \citenamefont {Barbiero}}]{macri_bound_2021}%
  \BibitemOpen
  \bibfield  {author} {\bibinfo {author} {\bibfnamefont {T.}~\bibnamefont
  {Macrì}}, \bibinfo {author} {\bibfnamefont {L.}~\bibnamefont {Lepori}},
  \bibinfo {author} {\bibfnamefont {G.}~\bibnamefont {Pagano}}, \bibinfo
  {author} {\bibfnamefont {M.}~\bibnamefont {Lewenstein}}, \ and\ \bibinfo
  {author} {\bibfnamefont {L.}~\bibnamefont {Barbiero}},\ }\bibfield  {title}
  {\enquote {\bibinfo {title} {Bound state dynamics in the long-range spin-1/2
  {XXZ} model},}\ }\href {\doibase 10.1103/PhysRevB.104.214309} {\bibfield
  {journal} {\bibinfo  {journal} {Phys. Rev. B}\ }\textbf {\bibinfo {volume}
  {104}},\ \bibinfo {pages} {214309} (\bibinfo {year} {2021})}\BibitemShut
  {NoStop}%
\bibitem [{\citenamefont {Runge}\ and\ \citenamefont
  {Zimanyi}(1994)}]{runge_exact-diagonalization_1994}%
  \BibitemOpen
  \bibfield  {author} {\bibinfo {author} {\bibfnamefont {K.~J.}\ \bibnamefont
  {Runge}}\ and\ \bibinfo {author} {\bibfnamefont {G.~T.}\ \bibnamefont
  {Zimanyi}},\ }\bibfield  {title} {\enquote {\bibinfo {title}
  {Exact-diagonalization study of the one-dimensional disordered \textit{{XXZ}}
  model},}\ }\href {\doibase 10.1103/PhysRevB.49.15212} {\bibfield  {journal}
  {\bibinfo  {journal} {Phys. Rev. B}\ }\textbf {\bibinfo {volume} {49}},\
  \bibinfo {pages} {15212--15222} (\bibinfo {year} {1994})}\BibitemShut
  {NoStop}%
\bibitem [{\citenamefont {Pal}\ \emph {et~al.}()\citenamefont {Pal},
  \citenamefont {Sharma}, \citenamefont {Changlani},\ and\ \citenamefont
  {Pujari}}]{pal_colorful_2021}%
  \BibitemOpen
  \bibfield  {author} {\bibinfo {author} {\bibfnamefont {S.}~\bibnamefont
  {Pal}}, \bibinfo {author} {\bibfnamefont {P.}~\bibnamefont {Sharma}},
  \bibinfo {author} {\bibfnamefont {H.~J.}\ \bibnamefont {Changlani}}, \ and\
  \bibinfo {author} {\bibfnamefont {S.}~\bibnamefont {Pujari}},\ }\bibfield
  {title} {\enquote {\bibinfo {title} {Colorful points in the {XY} regime of
  {XXZ} quantum magnets},}\ }\href {\doibase 10.1103/PhysRevB.103.144414} {\
  \textbf {\bibinfo {volume} {103}},\ \bibinfo {pages} {144414}}\BibitemShut
  {NoStop}%
\bibitem [{\citenamefont {Schollwöck}\ \emph {et~al.}(2008)\citenamefont
  {Schollwöck}, \citenamefont {Richter}, \citenamefont {Farnell},\ and\
  \citenamefont {Bishop}}]{schollwock_quantum_2004}%
  \BibitemOpen
  \bibfield  {author} {\bibinfo {author} {\bibfnamefont {U.}~\bibnamefont
  {Schollwöck}}, \bibinfo {author} {\bibfnamefont {J.}~\bibnamefont
  {Richter}}, \bibinfo {author} {\bibfnamefont {D.~J.~J.}\ \bibnamefont
  {Farnell}}, \ and\ \bibinfo {author} {\bibfnamefont {R.~F.}\ \bibnamefont
  {Bishop}},\ }\href {\doibase 10.1007/b96825} {\emph {\bibinfo {title}
  {Quantum {Magnetism}}}},\ Vol.\ \bibinfo {volume} {645}\ (\bibinfo
  {publisher} {Springer},\ \bibinfo {year} {2008})\BibitemShut {NoStop}%
\bibitem [{\citenamefont {Liu}\ \emph {et~al.}(2023)\citenamefont {Liu},
  \citenamefont {Gao}, \citenamefont {Chen}, \citenamefont {Henderson},
  \citenamefont {Dukelsky},\ and\ \citenamefont {Scuseria}}]{zhiyuan_arxiv}%
  \BibitemOpen
  \bibfield  {author} {\bibinfo {author} {\bibfnamefont {Z.}~\bibnamefont
  {Liu}}, \bibinfo {author} {\bibfnamefont {F.}~\bibnamefont {Gao}}, \bibinfo
  {author} {\bibfnamefont {G.~P.}\ \bibnamefont {Chen}}, \bibinfo {author}
  {\bibfnamefont {T.~M.}\ \bibnamefont {Henderson}}, \bibinfo {author}
  {\bibfnamefont {J.}~\bibnamefont {Dukelsky}}, \ and\ \bibinfo {author}
  {\bibfnamefont {G.~E.}\ \bibnamefont {Scuseria}},\ }\bibfield  {title}
  {\enquote {\bibinfo {title} {Exploring {Spin} {AGP} {Ansatze} for {Strongly}
  {Correlated} {Spin} {Systems}},}\ }\href {https://arxiv.org/abs/2303.04925}
  {\bibfield  {journal} {\bibinfo  {journal} {arXiv preprint arXiv:2303.04925}\
  } (\bibinfo {year} {2023})}\BibitemShut {NoStop}%
\bibitem [{\citenamefont {Majumdar}\ and\ \citenamefont
  {Ghosh}(1969)}]{majumdar_nextnearestneighbor_1969}%
  \BibitemOpen
  \bibfield  {author} {\bibinfo {author} {\bibfnamefont {C.~K.}\ \bibnamefont
  {Majumdar}}\ and\ \bibinfo {author} {\bibfnamefont {D.~K.}\ \bibnamefont
  {Ghosh}},\ }\bibfield  {title} {\enquote {\bibinfo {title} {On
  {Next}‐{Nearest}‐{Neighbor} {Interaction} in {Linear} {Chain}. {I}},}\
  }\href {\doibase 10.1063/1.1664978} {\bibfield  {journal} {\bibinfo
  {journal} {J. Math. Phys.}\ }\textbf {\bibinfo {volume} {10}},\ \bibinfo
  {pages} {1388--1398} (\bibinfo {year} {1969})}\BibitemShut {NoStop}%
\bibitem [{\citenamefont {Gelfand}(1990)}]{gelfand_series_1990}%
  \BibitemOpen
  \bibfield  {author} {\bibinfo {author} {\bibfnamefont {M.~P.}\ \bibnamefont
  {Gelfand}},\ }\bibfield  {title} {\enquote {\bibinfo {title} {Series
  investigations of magnetically disordered ground states in two-dimensional
  frustrated quantum antiferromagnets},}\ }\href {\doibase
  10.1103/PhysRevB.42.8206} {\bibfield  {journal} {\bibinfo  {journal} {Phys.
  Rev. B}\ }\textbf {\bibinfo {volume} {42}},\ \bibinfo {pages} {8206--8213}
  (\bibinfo {year} {1990})}\BibitemShut {NoStop}%
\bibitem [{\citenamefont {Zhitomirsky}\ and\ \citenamefont
  {Ueda}(1996)}]{zhitomirsky_valence-bond_1996}%
  \BibitemOpen
  \bibfield  {author} {\bibinfo {author} {\bibfnamefont {M.~E.}\ \bibnamefont
  {Zhitomirsky}}\ and\ \bibinfo {author} {\bibfnamefont {K.}~\bibnamefont
  {Ueda}},\ }\bibfield  {title} {\enquote {\bibinfo {title} {Valence-bond
  crystal phase of a frustrated spin-1/2 square-lattice antiferromagnet},}\
  }\href {\doibase 10.1103/PhysRevB.54.9007} {\bibfield  {journal} {\bibinfo
  {journal} {Phys. Rev. B}\ }\textbf {\bibinfo {volume} {54}},\ \bibinfo
  {pages} {9007--9010} (\bibinfo {year} {1996})}\BibitemShut {NoStop}%
\bibitem [{\citenamefont {Takano}\ \emph {et~al.}(2003)\citenamefont {Takano},
  \citenamefont {Kito}, \citenamefont {Ōno},\ and\ \citenamefont
  {Sano}}]{takano_nonlinear_2003}%
  \BibitemOpen
  \bibfield  {author} {\bibinfo {author} {\bibfnamefont {K.}~\bibnamefont
  {Takano}}, \bibinfo {author} {\bibfnamefont {Y.}~\bibnamefont {Kito}},
  \bibinfo {author} {\bibfnamefont {Y.}~\bibnamefont {Ōno}}, \ and\ \bibinfo
  {author} {\bibfnamefont {K.}~\bibnamefont {Sano}},\ }\bibfield  {title}
  {\enquote {\bibinfo {title} {Nonlinear sigma {Model} {Method} for the
  {J}$_{\textrm{1}}$-{J}$_{\textrm{2}}$ {Heisenberg} {Model}: {Disordered}
  {Ground} {State} with {Plaquette} {Symmetry}},}\ }\href {\doibase
  10.1103/PhysRevLett.91.197202} {\bibfield  {journal} {\bibinfo  {journal}
  {Phys. Rev. Lett.}\ }\textbf {\bibinfo {volume} {91}},\ \bibinfo {pages}
  {197202} (\bibinfo {year} {2003})}\BibitemShut {NoStop}%
\bibitem [{\citenamefont {Lante}\ and\ \citenamefont
  {Parola}(2006)}]{lante_ising_2006}%
  \BibitemOpen
  \bibfield  {author} {\bibinfo {author} {\bibfnamefont {V.}~\bibnamefont
  {Lante}}\ and\ \bibinfo {author} {\bibfnamefont {A.}~\bibnamefont {Parola}},\
  }\bibfield  {title} {\enquote {\bibinfo {title} {Ising phase in the
  {J}$_{\textrm{1}}$−{J}$_{\textrm{2}}$ {Heisenberg} model},}\ }\href
  {\doibase 10.1103/PhysRevB.73.094427} {\bibfield  {journal} {\bibinfo
  {journal} {Phys. Rev. B}\ }\textbf {\bibinfo {volume} {73}},\ \bibinfo
  {pages} {094427} (\bibinfo {year} {2006})}\BibitemShut {NoStop}%
\bibitem [{\citenamefont {Hu}\ \emph {et~al.}(2013)\citenamefont {Hu},
  \citenamefont {Becca}, \citenamefont {Parola},\ and\ \citenamefont
  {Sorella}}]{hu_direct_2013}%
  \BibitemOpen
  \bibfield  {author} {\bibinfo {author} {\bibfnamefont {W.-J.}\ \bibnamefont
  {Hu}}, \bibinfo {author} {\bibfnamefont {F.}~\bibnamefont {Becca}}, \bibinfo
  {author} {\bibfnamefont {A.}~\bibnamefont {Parola}}, \ and\ \bibinfo {author}
  {\bibfnamefont {S.}~\bibnamefont {Sorella}},\ }\bibfield  {title} {\enquote
  {\bibinfo {title} {Direct evidence for a gapless {Z} 2 spin liquid by
  frustrating {Néel} antiferromagnetism},}\ }\href {\doibase
  10.1103/PhysRevB.88.060402} {\bibfield  {journal} {\bibinfo  {journal} {Phys.
  Rev. B}\ }\textbf {\bibinfo {volume} {88}},\ \bibinfo {pages} {060402}
  (\bibinfo {year} {2013})}\BibitemShut {NoStop}%
\bibitem [{\citenamefont {Li}\ \emph {et~al.}(2012)\citenamefont {Li},
  \citenamefont {Becca}, \citenamefont {Hu},\ and\ \citenamefont
  {Sorella}}]{li_gapped_2012}%
  \BibitemOpen
  \bibfield  {author} {\bibinfo {author} {\bibfnamefont {T.}~\bibnamefont
  {Li}}, \bibinfo {author} {\bibfnamefont {F.}~\bibnamefont {Becca}}, \bibinfo
  {author} {\bibfnamefont {W.}~\bibnamefont {Hu}}, \ and\ \bibinfo {author}
  {\bibfnamefont {S.}~\bibnamefont {Sorella}},\ }\bibfield  {title} {\enquote
  {\bibinfo {title} {Gapped spin-liquid phase in the
  {J}$_{\textrm{1}}$−{J}$_{\textrm{2}}$ {Heisenberg} model by a bosonic
  resonating valence-bond ansatz},}\ }\href {\doibase
  10.1103/PhysRevB.86.075111} {\bibfield  {journal} {\bibinfo  {journal} {Phys.
  Rev. B}\ }\textbf {\bibinfo {volume} {86}},\ \bibinfo {pages} {075111}
  (\bibinfo {year} {2012})}\BibitemShut {NoStop}%
\bibitem [{\citenamefont {Nomura}\ and\ \citenamefont
  {Imada}(2021)}]{nomura_dirac-type_2021}%
  \BibitemOpen
  \bibfield  {author} {\bibinfo {author} {\bibfnamefont {Y.}~\bibnamefont
  {Nomura}}\ and\ \bibinfo {author} {\bibfnamefont {M.}~\bibnamefont {Imada}},\
  }\bibfield  {title} {\enquote {\bibinfo {title} {Dirac-{Type} {Nodal} {Spin}
  {Liquid} {Revealed} by {Refined} {Quantum} {Many}-{Body} {Solver} {Using}
  {Neural}-{Network} {Wave} {Function}, {Correlation} {Ratio}, and {Level}
  {Spectroscopy}},}\ }\href {\doibase 10.1103/PhysRevX.11.031034} {\bibfield
  {journal} {\bibinfo  {journal} {Phys. Rev. X}\ }\textbf {\bibinfo {volume}
  {11}},\ \bibinfo {pages} {031034} (\bibinfo {year} {2021})}\BibitemShut
  {NoStop}%
\bibitem [{\citenamefont {Roth}, \citenamefont {Szabó},\ and\ \citenamefont
  {MacDonald}(2022)}]{roth_high-accuracy_2022}%
  \BibitemOpen
  \bibfield  {author} {\bibinfo {author} {\bibfnamefont {C.}~\bibnamefont
  {Roth}}, \bibinfo {author} {\bibfnamefont {A.}~\bibnamefont {Szabó}}, \ and\
  \bibinfo {author} {\bibfnamefont {A.}~\bibnamefont {MacDonald}},\ }\bibfield
  {title} {\enquote {\bibinfo {title} {High-accuracy variational {Monte}
  {Carlo} for frustrated magnets with deep neural networks},}\ }\href
  {http://arxiv.org/abs/2211.07749} {\bibfield  {journal} {\bibinfo  {journal}
  {arXiv preprint arXiv:2211.07749}\ } (\bibinfo {year} {2022})}\BibitemShut
  {NoStop}%
\bibitem [{\citenamefont {Mayer}(1983)}]{mayer_behaviour_1983}%
  \BibitemOpen
  \bibfield  {author} {\bibinfo {author} {\bibfnamefont {I.}~\bibnamefont
  {Mayer}},\ }\bibfield  {title} {\enquote {\bibinfo {title} {On the behaviour
  of the {UHF} method near the “critical point”},}\ }\href {\doibase
  10.1007/BF03053757} {\bibfield  {journal} {\bibinfo  {journal} {Acta phys.
  Hung.}\ }\textbf {\bibinfo {volume} {54}},\ \bibinfo {pages} {249--266}
  (\bibinfo {year} {1983})}\BibitemShut {NoStop}%
\bibitem [{\citenamefont {Koga}, \citenamefont {Yamashita},\ and\ \citenamefont
  {Morokuma}(1991)}]{koga_incorrect_1991}%
  \BibitemOpen
  \bibfield  {author} {\bibinfo {author} {\bibfnamefont {N.}~\bibnamefont
  {Koga}}, \bibinfo {author} {\bibfnamefont {K.}~\bibnamefont {Yamashita}}, \
  and\ \bibinfo {author} {\bibfnamefont {K.}~\bibnamefont {Morokuma}},\
  }\bibfield  {title} {\enquote {\bibinfo {title} {On incorrect behavior of
  single annihilation equations of spin-projected {UHF} and {UMP} energies},}\
  }\href {\doibase 10.1016/0009-2614(91)85137-L} {\bibfield  {journal}
  {\bibinfo  {journal} {Chem. Phys. Lett.}\ }\textbf {\bibinfo {volume}
  {184}},\ \bibinfo {pages} {359--362} (\bibinfo {year} {1991})}\BibitemShut
  {NoStop}%
\bibitem [{\citenamefont {Tsuchimochi}\ and\ \citenamefont
  {Scuseria}(2011)}]{tsuchimochi_constrained_2011}%
  \BibitemOpen
  \bibfield  {author} {\bibinfo {author} {\bibfnamefont {T.}~\bibnamefont
  {Tsuchimochi}}\ and\ \bibinfo {author} {\bibfnamefont {G.~E.}\ \bibnamefont
  {Scuseria}},\ }\bibfield  {title} {\enquote {\bibinfo {title} {Constrained
  active space unrestricted mean-field methods for controlling
  spin-contamination},}\ }\href {\doibase 10.1063/1.3549134} {\bibfield
  {journal} {\bibinfo  {journal} {J. Chem. Phys.}\ }\textbf {\bibinfo {volume}
  {134}},\ \bibinfo {pages} {064101} (\bibinfo {year} {2011})}\BibitemShut
  {NoStop}%
\bibitem [{\citenamefont {Fishman}, \citenamefont {White},\ and\ \citenamefont
  {Stoudenmire}(2022)}]{itensor}%
  \BibitemOpen
  \bibfield  {author} {\bibinfo {author} {\bibfnamefont {M.}~\bibnamefont
  {Fishman}}, \bibinfo {author} {\bibfnamefont {S.~R.}\ \bibnamefont {White}},
  \ and\ \bibinfo {author} {\bibfnamefont {E.~M.}\ \bibnamefont
  {Stoudenmire}},\ }\bibfield  {title} {\enquote {\bibinfo {title} {{The
  ITensor Software Library for Tensor Network Calculations}},}\ }\href
  {\doibase 10.21468/SciPostPhysCodeb.4} {\bibfield  {journal} {\bibinfo
  {journal} {SciPost Phys. Codebases}\ ,\ \bibinfo {pages} {4}} (\bibinfo
  {year} {2022})}\BibitemShut {NoStop}%
\bibitem [{\citenamefont {Atkinson}(1989)}]{atkinson_introduction_1989}%
  \BibitemOpen
  \bibfield  {author} {\bibinfo {author} {\bibfnamefont {K.~E.}\ \bibnamefont
  {Atkinson}},\ }\href@noop {} {\emph {\bibinfo {title} {An introduction to
  numerical analysis}}},\ \bibinfo {edition} {2nd}\ ed.\ (\bibinfo  {publisher}
  {Wiley},\ \bibinfo {address} {New York},\ \bibinfo {year} {1989})\BibitemShut
  {NoStop}%
\bibitem [{\citenamefont {Tsuchimochi}\ and\ \citenamefont
  {Ten-no}()}]{tsuchimochi_orbital-invariant_2018}%
  \BibitemOpen
  \bibfield  {author} {\bibinfo {author} {\bibfnamefont {T.}~\bibnamefont
  {Tsuchimochi}}\ and\ \bibinfo {author} {\bibfnamefont {S.~L.}\ \bibnamefont
  {Ten-no}},\ }\bibfield  {title} {\enquote {\bibinfo {title}
  {Orbital-invariant spin-extended approximate coupled-cluster for
  multi-reference systems},}\ }\href {\doibase 10.1063/1.5036542} {\bibfield
  {journal} {\bibinfo  {journal} {J. Chem. Phys.}\ }\textbf {\bibinfo {volume}
  {149}},\ \bibinfo {pages} {044109}}\BibitemShut {NoStop}%
\bibitem [{\citenamefont {Duguet}()}]{duguet_symmetry_2015}%
  \BibitemOpen
  \bibfield  {author} {\bibinfo {author} {\bibfnamefont {T.}~\bibnamefont
  {Duguet}},\ }\bibfield  {title} {\enquote {\bibinfo {title} {Symmetry broken
  and restored coupled-cluster theory: I. rotational symmetry and angular
  momentum},}\ }\href {\doibase 10.1088/0954-3899/42/2/025107} {\bibfield
  {journal} {\bibinfo  {journal} {J. Phys. G: Nucl. Part. Phys.}\ }\textbf
  {\bibinfo {volume} {42}},\ \bibinfo {pages} {025107}}\BibitemShut {NoStop}%
\bibitem [{\citenamefont {Song}, \citenamefont {Henderson},\ and\ \citenamefont
  {Scuseria}()}]{song_power_2022}%
  \BibitemOpen
  \bibfield  {author} {\bibinfo {author} {\bibfnamefont {R.}~\bibnamefont
  {Song}}, \bibinfo {author} {\bibfnamefont {T.~M.}\ \bibnamefont {Henderson}},
  \ and\ \bibinfo {author} {\bibfnamefont {G.~E.}\ \bibnamefont {Scuseria}},\
  }\bibfield  {title} {\enquote {\bibinfo {title} {A power series approximation
  in symmetry projected coupled cluster theory},}\ }\href {\doibase
  10.1063/5.0080165} {\bibfield  {journal} {\bibinfo  {journal} {J. Chem.
  Phys.}\ }\textbf {\bibinfo {volume} {156}},\ \bibinfo {pages}
  {104105}}\BibitemShut {NoStop}%
\bibitem [{\citenamefont {Khamoshi}\ \emph {et~al.}(2021)\citenamefont
  {Khamoshi}, \citenamefont {Chen}, \citenamefont {Henderson},\ and\
  \citenamefont {Scuseria}}]{khamoshi_exploring_2021}%
  \BibitemOpen
  \bibfield  {author} {\bibinfo {author} {\bibfnamefont {A.}~\bibnamefont
  {Khamoshi}}, \bibinfo {author} {\bibfnamefont {G.~P.}\ \bibnamefont {Chen}},
  \bibinfo {author} {\bibfnamefont {T.~M.}\ \bibnamefont {Henderson}}, \ and\
  \bibinfo {author} {\bibfnamefont {G.~E.}\ \bibnamefont {Scuseria}},\
  }\bibfield  {title} {\enquote {\bibinfo {title} {Exploring non-linear
  correlators on {AGP}},}\ }\href {\doibase 10.1063/5.0039618} {\bibfield
  {journal} {\bibinfo  {journal} {J. Chem. Phys.}\ }\textbf {\bibinfo {volume}
  {154}},\ \bibinfo {pages} {074113} (\bibinfo {year} {2021})}\BibitemShut
  {NoStop}%
\bibitem [{\citenamefont {Henderson}\ and\ \citenamefont
  {Scuseria}(2020)}]{henderson_correlating_2020}%
  \BibitemOpen
  \bibfield  {author} {\bibinfo {author} {\bibfnamefont {T.~M.}\ \bibnamefont
  {Henderson}}\ and\ \bibinfo {author} {\bibfnamefont {G.~E.}\ \bibnamefont
  {Scuseria}},\ }\bibfield  {title} {\enquote {\bibinfo {title} {Correlating
  the antisymmetrized geminal power wave function},}\ }\href {\doibase
  10.1063/5.0021144} {\bibfield  {journal} {\bibinfo  {journal} {J. Chem.
  Phys.}\ }\textbf {\bibinfo {volume} {153}},\ \bibinfo {pages} {084111}
  (\bibinfo {year} {2020})}\BibitemShut {NoStop}%
\bibitem [{\citenamefont {Shen}\ and\ \citenamefont
  {Li}(2009)}]{shen_block_2009}%
  \BibitemOpen
  \bibfield  {author} {\bibinfo {author} {\bibfnamefont {J.}~\bibnamefont
  {Shen}}\ and\ \bibinfo {author} {\bibfnamefont {S.}~\bibnamefont {Li}},\
  }\bibfield  {title} {\enquote {\bibinfo {title} {Block correlated coupled
  cluster method with the complete active-space self-consistent-field reference
  function: {Applications} for low-lying electronic excited states},}\ }\href
  {\doibase 10.1063/1.3256297} {\bibfield  {journal} {\bibinfo  {journal} {J.
  Chem. Phys.}\ }\textbf {\bibinfo {volume} {131}},\ \bibinfo {pages} {174101}
  (\bibinfo {year} {2009})}\BibitemShut {NoStop}%
\end{thebibliography}%

\end{document}